\documentclass[12pt,letterpaper]{article}
\usepackage[affil-it]{authblk}

\usepackage[hyphenbreaks]{breakurl}
\usepackage{booktabs}
\usepackage{stmaryrd}
\usepackage{cite}

\usepackage[margin=1 in]{geometry}
\usepackage[T1]{fontenc} 
\usepackage{amssymb, amsmath, amsthm, mathtools, floatflt, palatino, euler, epic, eepic, microtype}
\usepackage{comment}
\usepackage{cases}
\usepackage{graphicx}
\usepackage[section]{placeins}
\usepackage{float}
\usepackage{booktabs}
\usepackage{algorithm}
\usepackage{algpseudocode}
\usepackage[normalem]{ulem}
\usepackage{hyperref}

\usepackage{multirow}
\usepackage[usenames, table, xcdraw]{xcolor}
\usepackage{subcaption}
\usepackage[skip=0.333\baselineskip]{subcaption}
\makeatletter
\let\UrlSpecialsOld\UrlSpecials
\def\UrlSpecials{\UrlSpecialsOld\do\/{\Url@slash}\do\_{\Url@underscore}}%
\def\Url@slash{\@ifnextchar/{\kern-.11em\mathchar47\kern-.2em}%
    {\kern-.0em\mathchar47\kern-.08em\penalty\UrlBigBreakPenalty}}
\def\Url@underscore{\nfss@text{\leavevmode \kern.06em\vbox{\hrule\@width.3em}}}
\makeatother

\hypersetup{
colorlinks=true,
linkcolor=black,
citecolor=black,
urlcolor=black
} 

\begin{document}

\title{\LARGE Use of mobile phone sensing data to estimate residence and occupation times in patches: Human mobility restrictions and COVID-19}

\author[1]{L. Leticia Ram\'irez-Ram\'irez} \affil[1]{Probability and Statistics, Centro de Investigaci\'on en Matem\'aticas A.C., Mexico}
\author[2]{Jos\'e A. Montoya} \affil[2]{Department of Mathematics, University of Sonora, Mexico}
\author[2]{Jes\'us F. Espinoza}
\author[3]{Chahak Mehta} \affil[3]{ Department of Aerospace Engineering and Engineering Mechanics, University of Texas at Austin, United States}
\author[1]{Albert Orwa Akuno} 
\author[3]{Tan Bui-Thanh}
\maketitle


\begin{abstract}

Modeling critical social phenomena, such as economic trends and infectious disease transmission, often requires capturing the dynamics of population mobility. This study focuses on modeling and inferring urban population mobility using geospatial data obtained from inhabitants' GPS reports. We estimate mobility patterns and the time fractions that inhabitants spend in each of the areas of interest, such as zip codes and census geographical areas, utilizing a Brownian bridge model. The derived information can be applied to diverse models involving human activities and dynamics. However, our primary objective is to address the practical gap in epidemic modeling based on patches, which may require more detailed mobility information than conventional origin-destination matrices provide. In practice, such information is often reduced to simpler structures or mobility models with few parameters that synthesize human mobility. We illustrate the model and method using data from the city of Hermosillo, Sonora, Mexico, in 2020, specifically between the two local waves of the 2019 coronavirus disease pandemic. We obtain estimations for different time periods to assess their stability and sensitivity, and compare these findings against known mobility restrictions and social events. Furthermore, we integrate the estimated residence and occupation parameters into a multi-patch compartmental epidemiological model to assess the impact on the epidemic evolution of changes in urban mobility.
\end{abstract}


\section{Introduction}

It is increasingly evident that we are witnessing a continual rise in global and regional connectivity. Understanding mobility patterns is crucial to developing more realistic mathematical models to comprehend various phenomena such as economic trends, patterns of violence, information dissemination, and the spread of infectious diseases \cite{kraemer2019utilizing, pucci2015mapping, Tocto2021AMM}. To practically utilize these more realistic mathematical models, we need to estimate their complex high-dimensional parameters. However, this estimation problem is often not addressed, as many works proposing new models tend to study them theoretically or through simulations. Alternatively, they may reduce complexity by employing specific mobility models, thereby reducing the dimensionality of the problem. For example, some dynamic human systems use mobility parameters described by gravity or radiation models, producing more specific mobility parameters like origin-destination mobility matrices.

In this study, we leverage big data from smartphone geolocation information and the Brownian Bridge model to propose estimating mobility parameters beyond the conventional origin-destination matrices used in network models. Specifically, we focus on what we term the Residence and Occupation Matrix (ROM). Unlike the origin-destination matrix, which is typically used for commuter and migrant models, the ROM does not describe the number of travels between any two zones, but instead depicts the expected fraction of time that an individual residing in a specific zone spends in all other zones. This matrix allows us to characterize mobility patterns based on the most significant zones for individuals, according to the time they spend in them during various activities such as work, school, and commuting.

Although human mobility and COVID transmission have been extensively researched in the field of GIScience, valuable insights for future research are provided by Zhang et al. \cite{zhang2022human}. These insights include 1. Fostering multidisciplinary collaboration, 2. enhancing mathematical models for analysis and prediction of disease transmission, simulation, and prediction, and 3. diversifying sources of mobility data to ensure accuracy and suitability.

This study addresses the first point as part of a multidisciplinary effort involving industry, the healthcare sector, epidemiologists, mathematicians, statisticians, and computer scientists. The complexity of data gathering, model development, statistical inference, and big data challenges necessitated this multidisciplinary approach. The outcomes of this research are directly related to \cite{akuno2023multi} and \cite{akuno2023inference}. Regarding the second point, the works cited address this aspect, since we propose the epidemic model and theoretically studied it in \cite{akuno2023multi}, and divide the estimation problem into two parts: first, the estimation of mobility parameters (presented here), and second, the statistical inference of remaining epidemic parameters \cite{akuno2023inference}. The inference of the epidemic model using this strategy is based on smartphone information and COVID-19 incidence data in Hermosillo.

Regarding the last point, we acknowledge that while other studies utilize similar geolocation (GPS) data, they often do not fully exploit its potential. Typically, these studies focus on estimating mobility information related to residence location and origin-destination matrices, relying solely on the frequency of GPS reports in each zone. In contrast, our utilization of the Brownian model enables us to estimate comprehensive spatio-temporal information from discrete-time geolocation reports, offering significant opportunities for modeling various human activities in a metropolitan area in detail. Moreover, our approach facilitates the incorporation of estimates derived from real-time or near-real-time data acquisition, holding considerable promise for enhancing the application of mathematical models. This has the potential to advance our understanding and response to social, economic, or health-related events, such as the epidemic model outlined in this study.

The structure of the paper is as follows: Section 2 offers background information on mobility studies and epidemic models that incorporate mobility components. Section 3 outlines the demographic and GPS data collection from mobile phones. In Section 4, we detail the methods for residence selection and introduce the Brownian Bridge model, which we employ to estimate occupation times. Section 5 presents the results of the residence selection and the estimated ROMs for the designated periods. Here, we compare the ROMs to assess their stability and sensitivity in light of epidemic data, mitigation strategies, and celebrations. Additionally, we examine the sensitivity of the epidemic model under realistic mobility scenarios based on estimated mobility parameters. Finally, Section 6 provides a discussion of the results and outlines avenues for future research.

\section{Background}


Detailed human mobility records of a particular regional population can be obtained from mobile phone records \cite{song2010limits} and such information can be used to predict future user's locations, validate human mobility models \cite{nguyen2012using}, study the role of human mobility and estimate the main characteristics of social networks \cite{wang2011human}, among others.
Indeed, the desire to understand and predict how the human population moves in the real world has been one of the main objectives of data collected by mobile phone service providers \cite{ponieman2013human}.
The diversity of applications of the study of human mobility cannot be over-emphasized. For instance, in the area of transportation, the study of human mobility can be used as a guide for the optimization of road networks and public transport systems, leading to efficient urban planning and engineering. Some other applications include humanitarian relief, health services, and identification of social graphs \cite{ponieman2013human, lu2012predictability,ruiz2021role,yan2021measuring, ren2021appropriate,newman2004finding,traag2019louvain}. 



The heterogeneous human interactions can significantly impact the dynamics we study. In epidemiology, traditional mass-action compartmental models have proven their usefulness for different infectious agents in modeling infection dynamics at the regional level. However, for several diseases, it is crucial to consider the structure of sub-populations (defined by zones or patches) and their more specific contacts patterns that are usually induced by the spatial configuration and mobility.

To introduce heterogeneous human contacts, epidemics in networks model for individual connections (or reaction-diffusion models on meta-population networks) have played an essential role in journals with a physics focus \cite{cota2021infectious,colizza2007reaction, wang2020network,siegenfeld2020impact}. Nevertheless, the model inference, fit or selection, uses highly computing-intensive numeric methods such as Markov chain Monte Carlo (MCMC) or particle filtering based on Monte Carlo simulations. Unfortunately, the scaling of these methods makes them prohibitive when considering medium- to larger-size populations. This paper focuses on estimating the information used in the class of patchy epidemic models \cite{bichara2018multi, akuno2023multi} that fall into an intermediate complexity between mass-action and networks models (at individual level). These models introduce population heterogeneities at the level of sub-populations and can include the within- and between-contact patterns induced by mobility. 

Since this work emphasizes the applications for patchy epidemic models, from now on we refer as ``patches'' to the geographical zones that partition the area of interest.

Several epidemic models have proposed the introduction of regional mobility patterns \cite{arino2003multi, bichara2015sis, bichara2018multi, espinoza2020mobility, yin2020novel, marquez2022multi}  for human-to-human infectious diseases \cite{moreno2017role, bernardo2021patterns, laxminarayan2020epidemiology} or vector-borne diseases \cite{liyanage2021impact, de2021effect, al2021impact, marshall2018mathematical}. These can incorporate mobility with an emphasis on a regional scale, which is critical in the increasing number of cases at the beginning of the outbreak \cite{meekan2017ecology}. However, we want to evaluate mobility on the scale of the metropolitan area since most of these movements occur in the context of the daily existence of 55\% of the world population living in cities \cite{zlotnik2017world, UnProspects}.

Numerous authors have sought to identify and model mobility patterns using a wide array of data sources, ranging from traditional sources such as surveys and census data to more contemporary sources like GPS data, social media feeds, and online platforms \cite{barbosa2018human}. While official data and social networks may offer insights into mobility at a macroscopic level, covering large regions such as states or countries, they often lack the granularity needed to understand mobility within a metropolitan area in detail, both spatially and temporally.

For decades, scientists have studied animal movement using tracking devices affixed to wildlife, which generate detailed datasets capturing  movement patterns within specific populations. More recently, geolocated smartphones have emerged as powerful tool for human tracking devices for humans, owing to their widespread adoption among a significant portion of the population.  This ubiquity facilitates the collection of vast geolocation datasets, which provide detailed insights into human mobility and behavior with unprecedented spatial and temporal resolution. Consequently, this enables the study of human travel patterns at finer scales in both time and space \cite{palmer2013new, pucci2015mapping, sakarovitch2018estimating, gariazzo2019multi}. Mobile phones represent a valuable source of information for studying various aspect of human behavior, environmental monitoring, transportation, social services, businesses \cite{furletti2014use}, and hold untapped potential as tools for public health.

While the potential utility of data sourced from mobile phones has been recognized for several years, analyses typically center around estimating population sizes in specific areas, identifying city activities, hotspots, and characterizing mobility patterns through contact networks \cite{calabrese2014urban} or metrics \cite{kishore2020measuring}. These insights are often derived directly from reported GPS records and aggregated in time intervals.

In the context of epidemiological models, researchers have focused on understanding certain types of movement that exhibit regular patterns, such as commuting between two regions \cite{tizzoni2014use}. Typically, these movements are estimated based on the frequency of GPS reports in each area. In more theoretical approaches, authors often assume specific models for human mobility from one region to another, such as adjacency and gravity models \cite{kraemer2019utilizing, barbosa2018human}. Additionally, the impact of mobility, according to these models, is illustrated under various mobility scenarios \cite{Tocto2021AMM}.

For these these epidemic models, the mobility estimation reduces to inferring the percentage of individuals for each pair of residence and destination regions. However, epidemiological models, such as \cite{bichara2018multi,akuno2023multi}, consider mobility not only as commuting between two patches, but also as the different temporal contacts that individuals have with individuals on routes to various destinations during the day. These other epidemiological models then use the information on the origin (residence) and the proportion of time individuals spend in each region (including their own). We propose estimating this Residence and Occupation Matrix (ROM) with a model that takes into consideration the individuals' sequence of GPS reports (`pings') and the stochastic variation from the path between consecutive locations.

The objective is to determine the residential patch of each individual and statistically estimate the average time spent by inhabitants in each of the $n$ patches of interest within the city, such as zip codes and other geographical areas. This involves generating an $n \times n$ non-negative matrix that delineates the proportion of time that individuals residing in a specific area are expected to spend in each patch. Our proposed method is based on the Brownian bridge process commonly employed in ecology to analyze animal movement. By leveraging this method, we can estimate an individual's location at any given time between successive pings. We applied this approach to data from Hermosillo, Mexico, during the COVID-19 pandemic. Additionally, we integrated the estimated residence-mobility matrix into a multi-patch compartmental SEIR model to evaluate the epidemic impact of mobility changes resulting from governmental interventions.

\section{Data}
\subsection{Geographic and demographic information}

\begin{figure}[htb]
	\centering
	\begin{subfigure}[c]{.45\textwidth} 
		\centering
		\includegraphics[width = \textwidth]{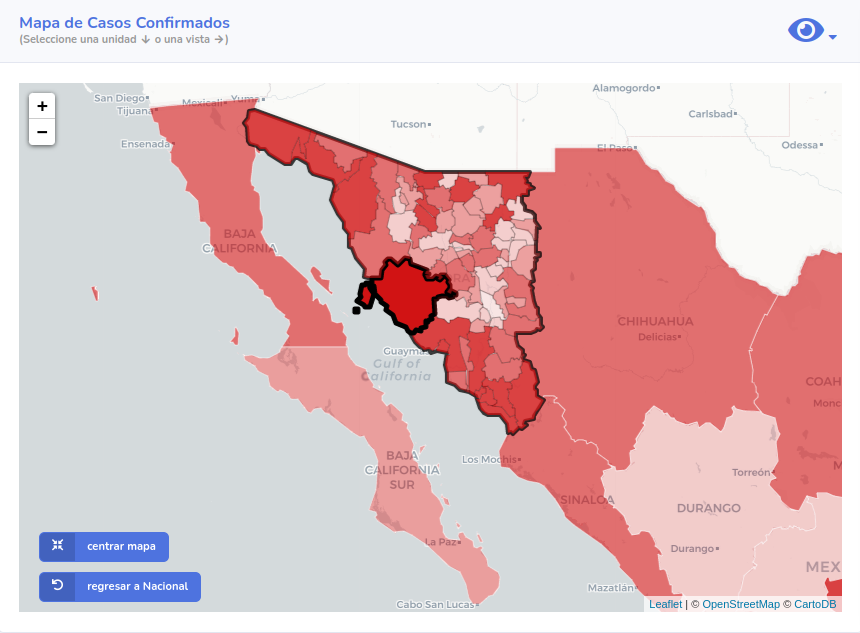}
		\caption{Hermosillo municipality, Sonora, Mexico \cite{covid19mexico}.}
		\label{fig:Hermosillo}
	\end{subfigure}
	\hfill
	\begin{subfigure}[c]{.5\textwidth} 
		\centering
		\includegraphics[width=\textwidth]{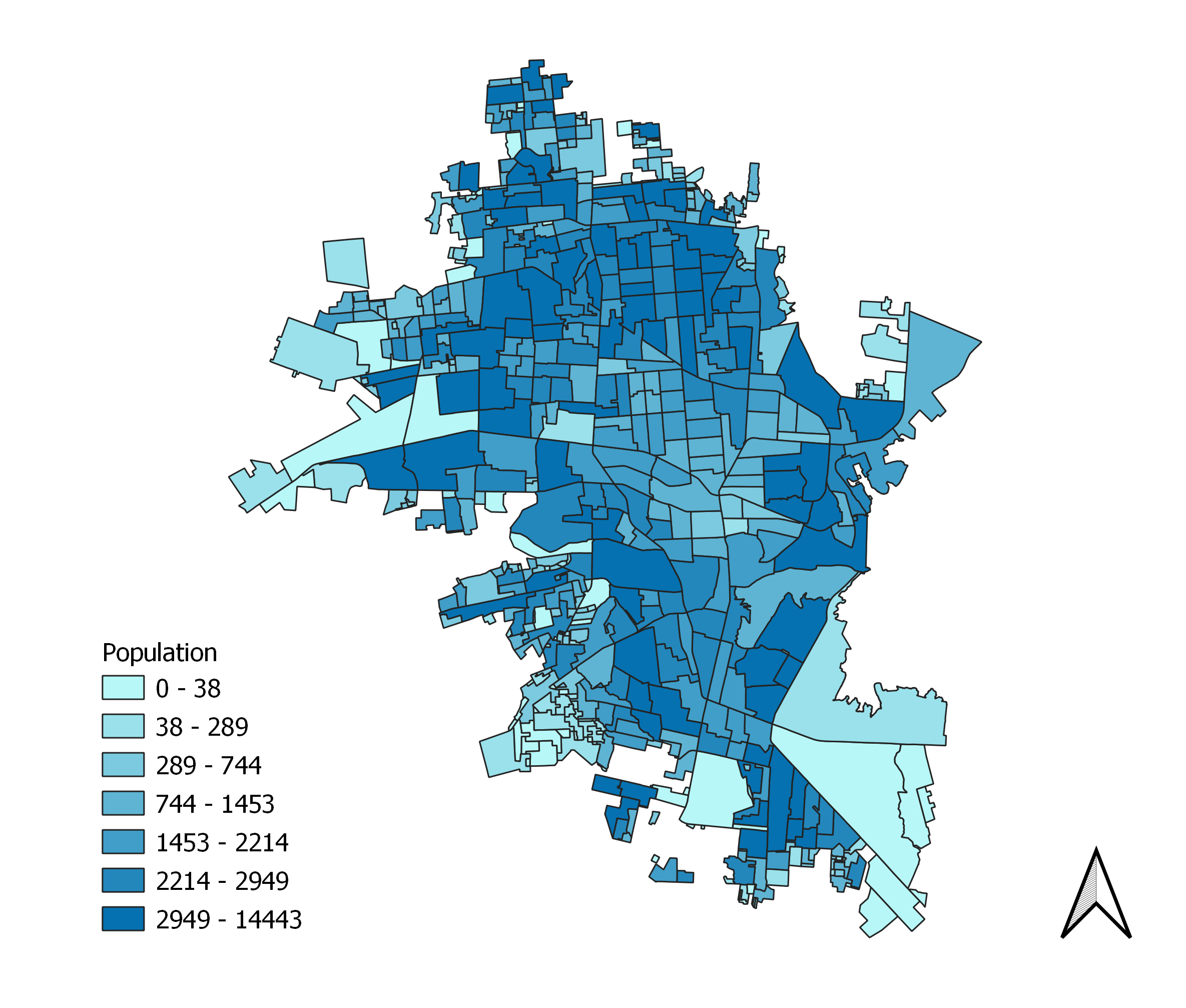}
		\caption{Hermosillo city and its population by AGEB \cite{inegi:shp_file}.}
		\label{fig:AGEB}
	\end{subfigure}
\caption{Hermosillo and its Basic Census Geographical Units (AGEBs). 
}
\label{fig:Hermosillo_AGEB}
\end{figure}

The city of Hermosillo is in the state of Sonora in the northwest of Mexico and has an area of 169.55 km$^2$. It is the capital of the state, and it is located 280 km south of the border with the United States and about 110 km from the coast of the Gulf of California. More precisely, it is located at the $29^\circ 05'$ parallel of the north latitude and the $110^\circ 57'$ meridian of west longitude from Greenwich (Figure~\ref{fig:Hermosillo}). The Population and Housing Census\footnote{Census 2020,  \url{https://www.inegi.org.mx/programas/ccpv/2020/}}, conducted by the National Institute of Statistics and Geography (INEGI) in 2020, reports a population of 2,944,840 in the state of Sonora, and the municipality with the largest population in this state corresponds to the city of Hermosillo, with a total population of 936,263. The local time zone in Hermosillo corresponds to UTC-07:00 throughout the year, and in contrast to all other states in Mexico (except  Quintana Roo), Sonora do not switch to the national summertime that in 2020 was from April 5 to October 25.

The city and its metropolitan area are divided into 502 basic census geographical units (AGEB), as shown in Figure~\ref{fig:Hermosillo_AGEB}. These areas represent the smallest demographic units, and information such as population and basic economics is publicly available for them. AGEBs, or aggregations thereof, present a natural option for defining unit areas (patches) for urban mobility analysis.

\subsection{Mobile phone sensing data}
Mobile phone tracking is a process to identify the location of a mobile phone based on various technologies and methods. These include multilateration (triangulation) of radio signals and using a global navigation satellite system (GNNS or GPS). The first technology uses the constant roaming radio signals that a mobile phone emits and may be picked up by three or more stationary cell towers enabling one to approximate the device location through triangulation. 
In contrast, the Global Positioning System (GPS, GNNS) allows for a high-precision localization (within a few centimeters to meters) of signals that operate independently of any telephonic or internet reception. 
A mobile phone `ping' is the process of sending a signal to a particular mobile phone and this responding with the request data to determine its location by utilizing its GPS location capabilities. 

The data that we use in this research corresponds to GPS pings collected by mobile service providers when users access specific (but undisclosed) apps for which they have granted permission to access their location information. 


On September 14, 2020, in the city of Hermosillo, Sonora, Mexico, the University of Sonora entered into an agreement granting access to GPS data from mobile phone reports covering the city area. This agreement, subject to confidentiality conditions, enables us to disseminate scientific findings derived from this data.

The raw data 
was weekly stored in a daily-partitioned table in BigQuery (a Google Cloud service). Through the API service and SQL queries, we could download specific sub-data bases as well as the whole database, in order to do the data wrangling prior to the analysis.

\subsubsection{The variables of mobility data}
The mobility data used in this work consists of 80,582,452 records, which contain the variables in Table \ref{TableVariables} that we describe afterwards.

\begin{table}
    \centering
    {\rowcolors{3}{lightgray!20}{white}
    \begin{tabular}{p{3.5cm}p{7cm}}
        \toprule
        Variable & Description \\
        \midrule
        \texttt{id\_adv} & mobile phone's ID (unique to each device) \\
        \texttt{timestamp}  & Ping date and time \\
        \texttt{lat} & Ping's Latitude\\
        \texttt{lon} & Ping's Longitude \\
    \end{tabular}}
    \caption{Variables in mobile phone sensing dataset.}
    \label{TableVariables}
\end{table}

\begin{description}
\item[\textbf{\texttt{id\_adv}}] It is a unique alphanumeric ID associated with each device (mobile phone). The database contains 306,963 such IDs. Each time a device utilizes a specific app with GPS localization capabilities enabled, it generates a ping, resulting in the addition of a new record to the database. The number of pings per device ranges from 1 to 18,297. However, 96.41\% of these devices have at most 1,000 records in the complete database. \\

\item[\textbf{\texttt{timestamp}}] This field consists of a sequence of characters in the format YYYY-MM-DD hh:mm:ss UTC, indicating the date and time when a device generated a ping. All timestamps fall within the timeframe from 00:00:00 on September 18, 2020, to 23:59:59 on December 13, 2020, in Universal Time Coordinated (UTC). However, for estimating urban mobility, conducting various sanity checks, and determining the residence patch for individuals in the city of Hermosillo (as outlined in Section \ref{subsec:Residence_selection}), we convert the timestamp variable to the local time zone. Therefore, the local dates and times of each mobility data observation range from 15:00 on September 17, 2020, to 15:00 on December 13, 2020. \\

\item[\textbf{\texttt{lat}, \texttt{lon}}] These two columns in the database contain numerical values indicating latitude and longitude coordinates of the device when a ping was made. To facilitate distance calculations between GPS data points, we transform these coordinate data into Cartesian coordinates using the UTM (Mercator) projection. This transformation was performed using the \texttt{geopandas} library in Python. The latitude values in the database range from 28.9753 to 29.1960, while the longitude values range from -111.1002 to -110.8457.




\end{description}


\subsubsection{Spatial and temporal characteristics of the data}
\begin{figure}[htb]
    \centering
    \includegraphics[width = \textwidth]{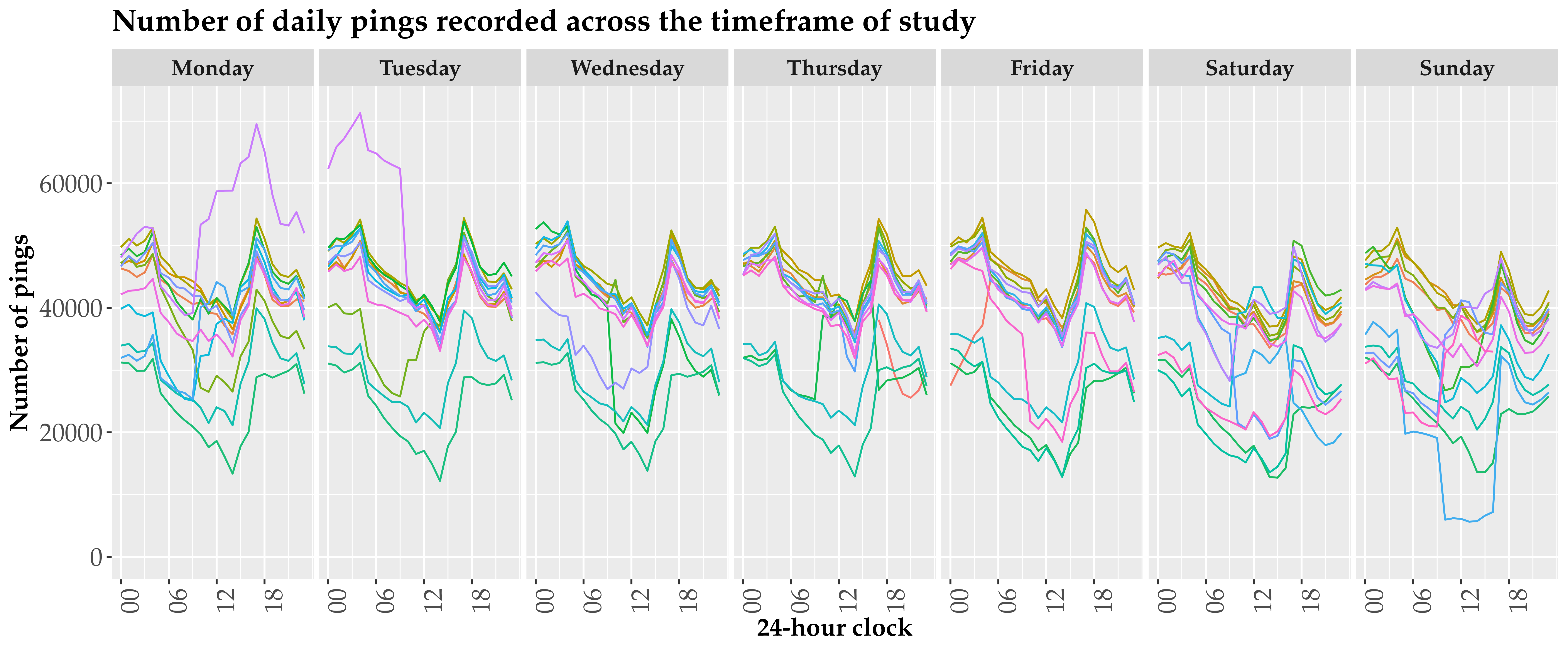}
    \caption{Number of pings by weekday and hour.}
    \label{fig:dayweek}
\end{figure}

Periodic patterns resulting from daily routines such as commuting, school, rush hours, meal times, and nighttime activities significantly impact the mobility of urban populations and the frequency of ping registrations. These patterns can vary based on factors like the day of the week (weekday or weekend) and the availability of local transport systems \cite{gariazzo2019multi}. In Figure~\ref{fig:dayweek}, we present  the complete time series of the number of pings categorized by both the day of the week and the time of day. This dataset spans 87 consecutive days from 2020-09-17 to 2020-12-13, and is divided in the plot into 13 weeks, starting from 00:00, 2020-09-14 (Monday) and ending on 24:00, 2020-12-13 (Sunday). Each week is visually differentiated by using a different color. 
	The first day (2020-09-17) is represented by the 'short' red line within the Thursday window. The conclusion of the time series is indicated by the 'short' pink line, which ends at 17:00 hrs within the Sunday window.
	This visualization reveals the temporal regularity, aligning with the typical activity schedules for city of Hermosillo residents.
		For example, activity levels tend to decrease at dawn and rise during hours associated with work and leisure. Notably, distinct patterns emerge for weekends.

\begin{figure}[ht]
    \centering    \includegraphics[width=0.75\textwidth]{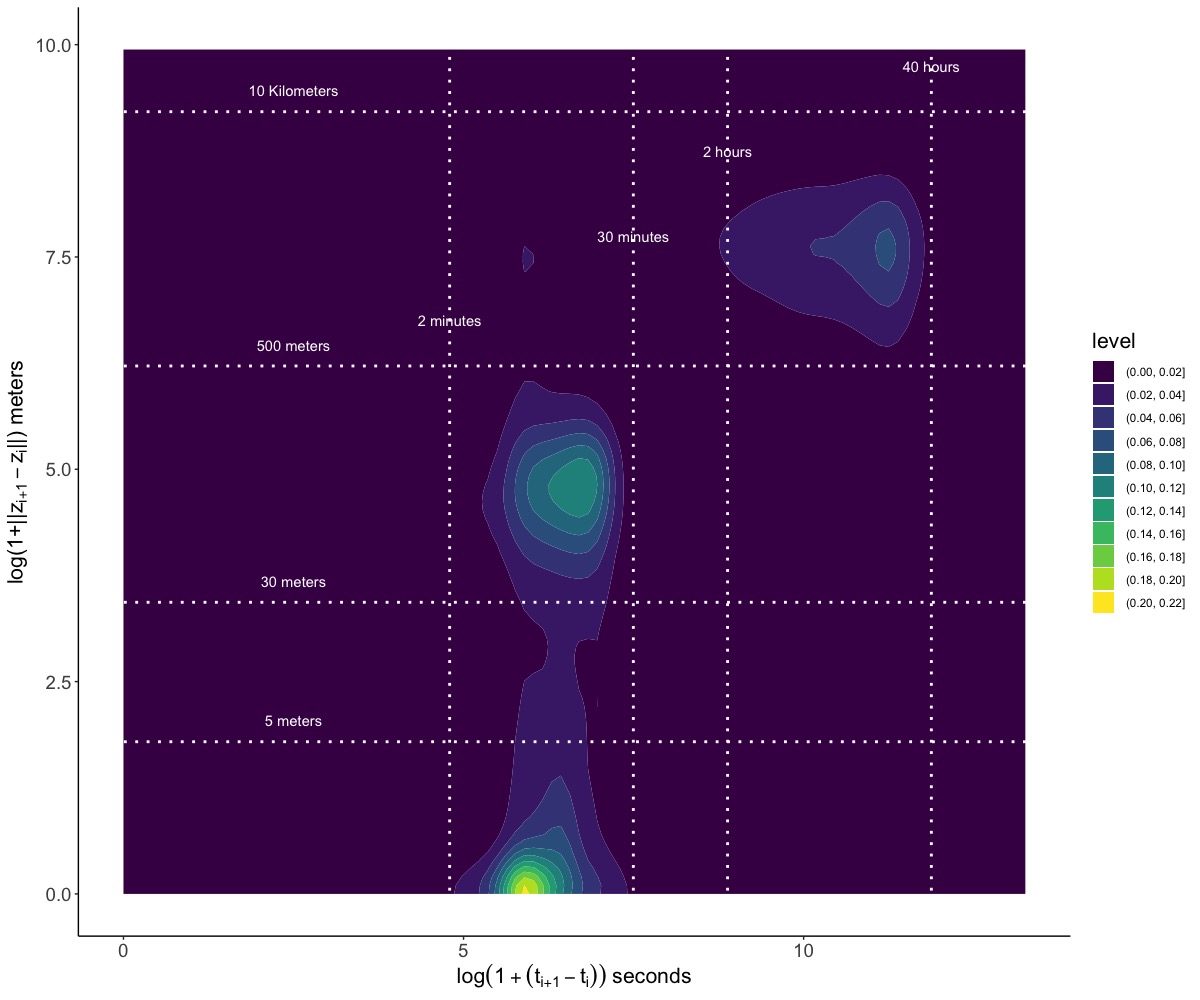}
    \caption{Kernel density of distance and consecutive pings.}
    \label{fig:DensitySample}
\end{figure}

Figure~\ref{fig:DensitySample} shows the bivariate normal kernel density estimate for traveled distance between consecutive pings and travel time (in a logarithmic scale for proper visualization) using all pings recorded during the study period from a sample of 10,000 id's. This graph was generated using MASS package in R and selecting kde2d's default parameters. The vertical dotted lines mark the corresponding time scale in seconds. Similarly, the horizontal dotted lines mark the corresponding distances traveled in meters. From the figure, we observe that most consecutive pings are at intervals between 2 and 30 minutes. These are divided into those where people move only few meters (likely walking or in a workspace) and those where they move between 30 and 500 meters (likely in a vehicle). At the top right, a third, less important group spends between 2 and 40 hours between pings but moves 500 to 10,000 meters.

To carry out censuses and surveys, it is necessary to define, in the geographical area, the study areas.
Such units are called Geostatistical Areas and in Mexico these are: State (AGEE), Municipal (AGEM) and Basic (AGEB). The latter is the smallest and fundamental geographical area and may be urban or rural. Urban AGEBs delimit a part or the total of a locality of 2500 inhabitants or more. In larger towns and cities, these AGEBs generally groups 25 to 50 blocks. In the case of the city of Hermosillo, there are 587 urban AGEBs whose population sizes are illustrated in Figure~\ref{fig:AGEB}.

\section{Methods}

\label{Sec:Brownian_Bridges}

One of the first papers that discussed urban analysis using data originated from mobile phones was \cite{ratti2006mobile} and was followed by works that highlighted possibilities of real-time visualization and monitoring of displacements in cities. In general terms, now we can distinguish models of human mobility aimed at reproducing individual mobility patterns or general population flows  \cite{barbosa2018human}. 

The residency-mobility matrix that we estimate can be viewed as a general population flow that describe the mobility patterns for individuals who inhabit each AGEB, but it is based on individual mobility modelling as we first estimate individual paths between consecutive pings. In the next sections we describe the model and the inference we propose.

\subsection{Residence selection}
\label{subsec:Residence_selection}
Since we are interested in describing mobility at the AGEB level, it is necessary to identify the AGEB to which each ping's location belongs. For this, it is essential to transform the latitude and longitude data to the UTM coordinate system and identify the AGEB to which it corresponds. We do this using the AGEBs official polygon information.

Now, as we also want to describe the mobility patterns for all individuals that live in each AGEB, and this information is not available for each ID, it is necessary to assign each ID to an AGEB as a residence using the very same database. That is, based on the ping information of each ID we determine the AGEB that its resides. For this, we use the an heuristic that combines criteria of frequency, timeframes, and AGEBs' population size. This selection process is described in Algorithm~\ref{alg:cap}. 

\begin{algorithm}
	\caption{Find the residence of all individuals}
	\label{alg:cap}
	\begin{algorithmic}
		\For{each individual $i$}
		\State $s1 \gets \text{AGEBs with maximum GPS data points}$
		\State $s2 \gets \text{AGEBs with maximum GPS data points between 22:00 and 06:00}$
		\State $f = s1 \cap s2$
		\If{len($f$) is 1}
		\State $\text{residence}_i \gets f$
		\Else
		\State $\text{residence}_i \gets f_i \text{ where $i$ is selected randomly weighted by population of $f_i$}$
		\EndIf
		\EndFor
	\end{algorithmic}
\end{algorithm}

The primary idea behind this algorithm is to determine the AGEB of residence for an ID based on where it demonstrates the highest frequency of pings between 22:00 and 06:00 hours. If multiple AGEBs fulfill this criterion, the residence is randomly assigned by sampling from these candidates, with the sampling weighted by their respective population sizes. This heuristic falls under the category of unsupervised methods and represents a variation of the "grid frequency method" \cite{zhao2022estimating,verma2024comparison}. The main distinctions lie in the fact that the grids correspond to irregularly shaped patches (AGEBs) that partition the area of interest, and it doesn't necessitate assigning a specific geolocation, such as a patch's centroid, but only the patch ID of residence.

\subsection{The Brownian bridge model}
Suppose that $Z_{rj}(t)$ represents the position of the \textit{j}-th resident from \textit{r}-th patch at time $t$. Then $r \in \{1,\ldots, n\}$, as $n$ is the number of considered patches,  and $j\in \{1,\ldots N_r\}$ where $N_{r}$ is the resident population size in patch $r$.

Let $\{z_{rj(1)},t_{rj(1)}\},\{z_{rj(2)},t_{rj(2)}\}$, \dots ,$\{z_{rj(n_{rj})},t_{rj(m_{rj})}\}$ be the information from the GPS reports for the $j$-th individual that resides in patch $r$. The term $m_{rj}$ corresponds to the total of pings  collected  during the observation time, and $z_{rj(i)}$ denotes the position of the $j$-th resident from the $r$-th patch at its $i$-th reporting time $t_{rj}(i)$. If we set the observation period starting with the first ping, then the individual observation period is $[0,T_{rj}]$ with $T_{rj}=t_{rj(n)}-t_{rj(1)}$. 

Now, let $Z^{[k]}_{rj}(t)$ denote the unobserved position of the \textit{j}-th resident from \textit{r}-th patch at time $t \in \left(t_{rj(k)},t_{rj(k+1)}\right)$ which undertakes a random walk from positions $z_{rj(k)}$ to $z_{rj(k+1)}$. 
If a Brownian motion is assumed between these two positions, then $Z^{[k]}_{rj}(t)$ is distributed according to a bivariate normal distribution at time $t \in \left[t_{rj(k)},t_{rj(k+1)}\right]$, with mean vector and covariance matrix given by
\begin{align}
    \mu(t;z_{rj(k)},z_{rj(k+1)}) &= z_{rj(k)}+(z_{rj(k+1)}-z_{rj(k)}) \left(\frac{t-t_{rj(k)}}{t_{rj(k+1)}-t_{rj(k)}}\right),\nonumber\\
    \Sigma(t;\sigma_{rj}) &=\sigma^{2}_{rj}\left[\frac{(t-t_{rj(k)})(t_{rj(k+1)}-t)}{(t_{rj(k+1)}-t_{rj(k)})}\right]\mathbb{I}, \nonumber
\end{align}
where $\mathbb{I}$ is the $2 \times 2$ identity matrix and $\sigma^{2}_{rj}$ is the Brownian motion variance related to the mobility of the $j$-th resident from $r$-th patch. Thus, the Brownian bridge process has the property of being normal along a straight line joining the points $z_{rj(k)}$ and $z_{rj(k+1)}$. In addition, the maximum variability is obtained at the middle of the trajectory  and the variance equals 0 when $t=t_{rj(k)}$ or $t=t_{rj(k+1)}$. This assumption is adequate when the location reports are precise, but in many applications it is convenient to introduce the uncertainty related to measurement errors. For this, we consider that the starting and ending locations are bivariate normal $\text{MVN}\left(z_{rj(k)},\delta^2\mathbb{I}\right)$  and $\text{MVN}\left(z_{rj(k+1)},\delta^2\mathbb{I}\right)$, where $\delta^2$ is the variance of the location error. 
Therefore the expected time spent at location $z$,  during $\left[t_{rj(k)},t_{rj(k+1)}\right]$, is
\begin{align}
    h^{[k]}_{rj}\left(z;\sigma_{rj},\delta\right)
    &= \frac{1}{T^{[k]}_{rj}} \int_{t_{rj(k)}}^{t_{rj(k+1)}} \phi \left(z; \mu^{[k]}_{rj}(t), \sigma^{[k]}_{rj}(t; \sigma_{rj}, \delta)\mathbb{I} \right) dt, 
    \label{MarginalDensity02}
\end{align}
where  $\phi\left(\cdotp;\mu,\Sigma\right)$ is the probability density function of a bivariate normal distribution with mean vector $\mu$ and covariance variance $\Sigma$, and 
\begin{align*}
    T^{[k]}_{rj} &= t_{rj(k+1)} - t_{rj(k)}, \\
    \mu^{[k]}_{rj}(t) &= z_{rj(k)} + \left(z_{rj(k+1)} - z_{rj(k)}\right) \alpha^{[k]}_{rj}(t), \\
    \sigma^{[k]}_{rj}(t; \sigma_{rj}, \delta) &= T^{[k]}_{rj} \alpha^{[k]}_{rj}(t)\left(1 - \alpha^{[k]}_{rj}(t)\right) \sigma^{2}_{rj} + \left(1 - \alpha^{[k]}_{rj}(t)\right)^2 \delta^2 + \left(\alpha^{[k]}_{rj}(t) \delta\right)^{2},\\
    \alpha^{[k]}_{rj}(t) &=\left( t- t_{rj(k)} \right)/T^{[k]}_{rj}.
\end{align*}

Using the $(m_{ri}-1)$ bridges, we then can estimate the density of the expected occupation time  for each individual, as a  mixture of the densities (\ref{MarginalDensity02}). That is 
\begin{equation}
h_{rj} \left(z; \sigma_{rj}, \delta \right) = \sum_{k=1}^{m_{rj}-1} \left(\frac{T^{[k]}_{rj}}{T_{rj}} \right) h^{[k]}_{rj} \left(z; \sigma_{rj}, \delta \right).
    \label{MarginalDensity03}
\end{equation}
For computing the overall fraction of time spent in a region $A$ (occupation time in $A$), by individuals from patch $r$, we average the individual fractions of times in $A$ considering all its patch residents as
\begin{equation}
\label{timePatch}
    P_{r} \left(A\right) =\frac{1}{N_{r}} \sum_{j=1}^{N_{r}}\int_A h_{rj} \left(z; \sigma_{rj}, \delta \right)dz. 
\end{equation}
For this work, we are interested in estimating $P_r(s)$ for each pair of patches $r$ and $s$ and these component constitute the residence and occupation matrix (ROM). If we have that individuals can travel only to any of the $n$ considered patches, the sum of the rows for the matrix $[P_{rs}=P_r(s)]$ will add up to one.

\subsection{Likelihood inference}
First, we assume that the variance of the location error $\delta^2$ can be known from the device specifications, while the individual variance $\sigma_{rj}$ of the mobility of the \textit{j}-th resident from $r$-th patch is unknown. We estimate $\sigma_{rj}$ using the method proposed in \cite{Horne} (also, see \cite{penunuri}). Therefore, we consider the $n_{rj}$ as odd numbers and estimate the mean based on the independent Brownian bridges for the non-overlapping time intervals
$$
\left\{\left[t_{rj(k-1)},t_{rj(k+1)}\right]\right\}_{k\in \{2,4,\ldots, m_{rj}-1\} }\ \ ,
$$ 
while regarding the in-between observation times $t_{rj(k)}$ as independent observations from these bridges to estimate $\sigma_{rj}$. Thus, $Z_{rj}(t_{rj(k)})$ is a  bivariate normal random variable with mean and covariance matrix given by
\begin{align}
    \mu(t_{rj(k)};z_{rj(k-1)},z_{rj(k+1)}) &= z_{rj(k-1)}+(z_{rj(k+1)}-z_{rj(k-1)})\alpha_{rj(k)} , \nonumber\\
	\Sigma(t_{rj(k)};\sigma_{rj}) &= \left[ T_{rj(k)} \alpha_{rj(k)}(1 - \alpha_{rj(k)}) \sigma^{2}_{rj} + (1 - \alpha_{rj(k)})^2 \delta^2 + \alpha_{rj(k)}^{2} \delta^{2} \right] \mathbb{I}, \nonumber
	\label{BBEstimation}
\end{align}
where 
$ T_{rj(k)} = t_{rj(k + 1)} - t_{rj(k - 1)}$ and $ \alpha_{rj(k)} = (t_{rj(k)} - t_{rj(k - 1)})/T_{rj(k)}$. 

Therefore, the likelihood function of the parameter $\sigma_{rj}$ can be written as
\begin{equation}
    L \left(\sigma_{rj}\right) = \displaystyle \prod_{k=
    2,4,\ldots, m_{rj}-1} \phi \left(z_{rj(k)}; \mu(t_{rj(k)}; z_{rj(k - 1)}, z_{rj(k + 1)}), \Sigma(t_{rj(k)}; \sigma_{rj}) \right),
    \label{LikelihoodFunction}
\end{equation}
where again $\phi$ denotes the probability density function of a bivariate normal distribution.

The maximum likelihood estimate for  $\sigma_{rj}$ is the value $\hat{\sigma}_{rj}$ that maximizes $L(\sigma_{rj})$ in (\ref{LikelihoodFunction}). Thus for any individual originating from patch $r$, the estimated probability density function at position $z$ is given by
\begin{equation*}
    \hat{h}^{*}_{r} \left(z\right) = \frac{1}{N_{r}} \sum_{j=1}^{N_{r}}h_{rj} \left(z; \hat{\sigma}_{rj}, \delta \right).
    \label{MarginalDensity06}
\end{equation*}
Then we can estimate the expected occupation time in $A$ from resident of patch $r$, based on (\ref{timePatch}), as

\begin{equation}
    \hat{P}_r(A)=\int_A\hat{h}^{*}_{r} \left(z\right)dz =\frac{1}{N_{r}}\sum_{j=1}^{N_{r}}\int_A h_{rj} \left(z; \hat{\sigma}_{rj}, \delta \right)dz. 
    \label{MarginalDensity05}
\end{equation}

If the information of the device's location error $\delta^2$ is unknown, the Markov property in the previous model is not longer true, but we can opt for the BMME extension proposed by \cite{pozdnyakov2014modeling} based on the joint distribution of $\left(B_{t}, Z_{rj}(t_{rj(1)}), \ldots, Z_{rj}(t_{rj(m_{rj})})\right)^{\top}$ where 
$Z_{rj}(t_{rj(i)})=B_{t_{rj(i)}}+\xi_i$, $i=1, \ldots, m_{rj}$,\   $\{\xi_i\}$ are iid Gaussian variables with mean 0 and variance $\delta^2$ that are also independent of the Brownian motion $\{B_{t}, t\geq 0\}$.

Since $\left(B_{t}, Z_{rj}(t_{rj(1)}), \ldots, Z_{rj}(t_{rj(m_{rj})})\right)^{\top}$ is a multivariate gaussian vector, then conditional ditribution 
$$
\operatorname{Pr}\left(B_t \in d x \mid \mathbf{Z}\right)
\sim \text{MVN}\left(\mu^{[k*]}_{rj}(t)=\Sigma_{12}\Sigma_Z^{-1}\boldsymbol{Z},\ \  \sigma^{[k*]}_{rj}(t)=\sigma_{rj}^2t-\Sigma_{12}\Sigma_Z^{-1}\Sigma_{12}^{\top}\right)
$$
where
\begin{align*}
	\boldsymbol{Z}&=\left(Z_{rj}(t_{rj(1)}),\ldots,Z_{rj}(t_{rj(m_{rj})})\right),\\
	[\Sigma_Z]_{i,s}&=\text{Cov}(Z_{rj}(t_{rj(i)}),Z_{rj}(t_{rj(s)}))=\left\{\begin{array}{ll}
		\sigma_{rj}^2 t_{rj(i)}+\delta^2 & \text{if } \ \ i=s\\[5pt]
		\sigma_{rj}^2 \min\left\{t_{rj(i)},t_{rj(s)} \right\}& \text{if } \ \  i\neq s\\
	\end{array}\right., \mbox{ and}\\
	\Sigma_{12}&=\sigma_{rj}^2\left[\min \left(t,t_{rj(1)}\right), \ldots, \min \left(t, t_{rj(m_{rj})}\right)\right]
\end{align*}

Then the density of the expected occupation time for each individual (\ref{MarginalDensity03}) under BMME, corresponds to
\begin{equation}
h_{rj}(z; \sigma_{rj},\delta)=\sum_{k=1}^{m_{r j}-1} \left(\frac{T_{r j}^{|k|}}{T_{r j}}\right)\left[
\frac{1}{T_{r j}^{|k|}} \int_{t_{rj(k)}}^{t_{rj(k+1)}} \phi \left(z; \mu^{[k*]}_{rj}(t), \sigma^{[k*]}_{rj}(t) \right) dt
\right].
\label{BMME}
\end{equation}

Similarly to the previous model, we can compute the maximum likelihood estimates for $\sigma_{rj}$ and $\delta$ using the likelihood function based on the joint density of the increments $x_{rj(t)}=z_{rj(t+1)}-z_{rj(t)}$
\begin{equation}
	L \left(\sigma_{rj},\delta\right) = 
		\phi \left((x_{rj(1)},x_{rj(2)},\ldots,x_{rj( m_{rj}-1)})^{\top}; \boldsymbol{0},  \Sigma_X\right),
	\label{LikelihoodFunction_BMME}
\end{equation}
where $\Sigma_X$ is the symmetric and banded matrix with $i,s$ elements
$$
	[\Sigma_X]_{i,s}=\left\{\begin{array}{ll}
	\sigma_{rj}^2 (t_{rj(i)}-t_{rj(i-1)})+2\delta^2 & \text{if } \ \ i=s\\[5pt]
	-\delta^2& \text{if } \ \  |i-s|=1\\[5pt]
	0& \text{if } \ \  |i-s|>1.\\
\end{array}\right.
$$
where $i,u \in\{1, 2, \ldots ,m_{r j}-1\}$.

The downside of introducing $\delta$ as a parameter extra to estimate is outweighed by the advantage of using all the data at the same time to estimate all parameters and the fact that this inference is not computationally more expensive to implement. Once the maximum likelihood estimates are calculated we obtain the expected occupation time in region $A$ similarity to (\ref{MarginalDensity05}) and using (\ref{BMME}),
\begin{equation}
	\hat{P}_r(A)=\frac{1}{N_{r}}\sum_{j=1}^{N_{r}}\int_A h_{rj} \left(z; \hat{\sigma}_{rj}, \hat{\delta} \right)dz. 
	\label{MarginalDensity_BMME}
\end{equation}
In general, the considered patches can have very irregular shapes and for computing the integral part in the previous expression we
require to resort to numerical methods. For the application we opt for the uniform sampling Monte Carlo integration. 

\section{Results}

\subsection {Data filtering and residence selection}
In this section, we describe the periods of time we consider for the ROMs estimations, the filtering process to select the devices to include in each estimation, and the resulting residence selection.


We have chosen three periods, each divided into two parts,  to estimate the mobility patterns. The objective is to compare the resulting estimates matrices pairwise for each period. Table \ref{table:periods_definitions} presents these time frames.

\begin{table}
\centering
\begin{tabular}{@{}llll@{}}
\toprule
\textbf{Abreviation} & \textbf{Full name}  & \textbf{Start date} & \textbf{End date}\\ \midrule
$P1_A$    & First period - First Part   & 2020-09-21  & 2020-10-04  \\
$P1_B$    & First period - Second Part   & 2020-10-26  & 2020-11-08  \\[4pt]
$P2_A$    & Second period - First Part   & 2020-09-21  & 2020-10-04  \\
$P2_B$    & Second period - Second Part   & 2020-11-02  & 2020-11-15  \\[4pt]
$P3_A$    & Third period - First Part   & 2020-09-21  & 2020-10-11  \\
$P3_B$    & Third period - Second Part   & 2020-10-12  & 2020-11-01  \\ \bottomrule
\end{tabular}
\caption{Periods within which the mobility matrices were estimated.}
\label{table:periods_definitions}
\end{table}

\begin{figure}[htb]
	\centering
	\includegraphics[width = \textwidth]{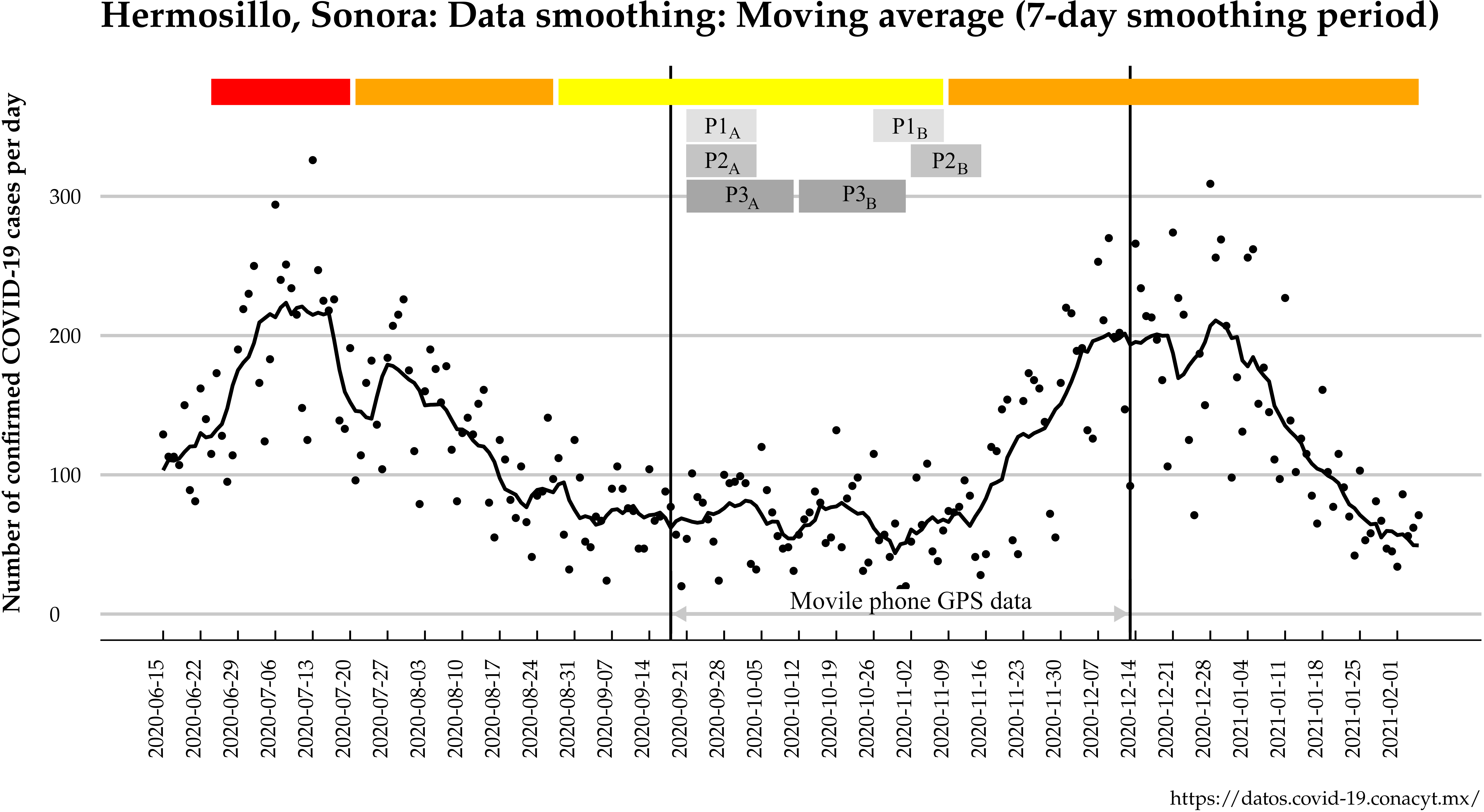}
	\caption{Daily confirmed cases of COVID-19 in Hermosillo, Sonora in the time frame of study. The points corresponds to the daily observations, the solid black to the smoothed epidemic curve. The  top colored bar indicates the local epidemic traffic light, and the grey indicate the periods and parts in Table~\ref{table:periods_definitions}.}
	\label{fig:Figure_SelectionPeriodsResidenceMatrix}
\end{figure}

The reason for selecting these periods was to study the moderate to possible important changes in mobility patters in relation to the implemented control measures for COVID-19.  The Secretary of Health of the State of Sonora adopted the ``epidemic traffic light'' and according to its colors (green, yellow, orange and red) the local government implemented specific measures. Some were recommendations, such as remote working and protective measures, but some other were more strictly imposed, such as limitations on gatherings, capacity restrictions for businesses and remote schooling. Figure \ref{fig:Figure_SelectionPeriodsResidenceMatrix} shows the number of daily  COVID-19 cases for the city of Hermosillo between the second half of 2020 and the beginning of 2021. The existing values for the epidemic traffic light are indicated on the top of this graph. 

To visualize the general epidemic pattern  Figure~\ref{fig:Figure_SelectionPeriodsResidenceMatrix} includes the moving average of cases. It also indicates the period of time for which we have the GPS information as the interval between the two vertical lines. Since the vaccination campaign started on 2021, all controls during this period were in relation to hygiene protocols, travel restriction, social distancing,  and stay-at-home orders.

The contemplated periods and its parts are also represented in Figure~\ref{fig:Figure_SelectionPeriodsResidenceMatrix} as the grey segments above the curve. It becomes evident that all selected periods primarily fall within the yellow alert phase for COVID-19 and are positioned between the fist and second epidemic waves in Hermosillo. Additionally, the second parts of periods 1 and 3 exhibit no indication of a significant rise in confirmed COVID-19 cases. In fact, the moving average shows a slight downward trend between 2020-09-21 and 2020-11-15. Therefore, with regard to confirmed cases, the second part of each period does not foreshadow the onset of the second wave of COVID-19 cases. 

Regarding the urban mobility, we postulate that the second part of each period could coincide with an increase in the movement of people within the city, as they encompass the dates of \textit{Halloween} (November 31st) as well as the traditional Mexican holiday \textit{D\'ia de Muertos} (November 1st and 2nd). Both festivities are fervently celebrated in cemeteries, markets, and streets. This surge in mobility may be associated to the delayed rise in COVID-19 cases that marked the beginning of the second wave.

\subsubsection{Data selection}
The original mobility data in the database consists of 80,582,452 pings that register the timestamp of 306,963 devices (IDs). However, to proceed with the estimation, we filter them to ensure a minimum of pings per device in each part. Table~\ref{table:Ids_union} contains the number of IDs that in any of the two parts,  of the corresponding period,  report at least 5, 11, 15, or 21 weekly pings. The table also shows its magnitude as the percentage of the number of IDs with at least one ping in any of both parts.

\begin{table}
\centering
\begin{tabular}{@{}lllll@{}}
	\toprule
	\textbf{Period}            & \textbf{+5 pings} & \textbf{+11 pings} & \textbf{+15 pings} & \textbf{+21 pings} \\ \midrule
	First & 217,396 (88.74\%)  & 153,117 (62.5\%)    & 85,742 (35\%)       & 44,859 (18.31\%)    \\
	Second & 226,174 (89.05\%)  & 154,374 (60.78\%)   & 89,831 (35.37\%)    & 46,456 (18.29\%)    \\
	Third & 186,448 (88.12\%)  & 128,915 (60.93\%)   & 68,485 (32.37\%)    & 32,416 (15.32\%)    \\ \bottomrule
\end{tabular}
\caption{IDs distribution according to their number of weekly pings per period.}
\label{table:Ids_union}
\end{table}

To perform the Brownian bridges estimation,  we filter the mobile phone database for each period by dropping the IDs with less than 10 pings in a week. That is, for each period and each part, we keep IDs with at least 11 weekly pings in any of the two parts. Then the number of selected IDs for each period-part are: 108,252 ($P1_A$), 123,878 ($P1_B$), 108,252 ($P2_A$),  120,156 ($P2_B$), 102,091 ($P3_A$), and  103,184 ($P3_B$). 

The population in Hermosillo city is about 936,263 inhabitants, then the number of IDs selected to represent the mobility is above 10.9\% for any period-part.




\subsection{Estimation of residence and occupation matrices for Hermosillo}
Based on the selected IDs and their pings information during each period-part (Table \ref{table:periods_definitions}), we obtained the maximum likelihood estimates for $\sigma_{rj}$, ($r=1,\ldots, n$, $j=1,\ldots, m_{rj}$) and $\delta$. 

With this estimates, in turn, we are able to approximate the  ROM for each period-part using (\ref{MarginalDensity_BMME}) as 

\begin{equation}
[P]_{ij}=\hat{P}_i(j)=\frac{1}{N_{i}}\sum_{k=1}^{N_{i}}\int_j h_{ik} \left(z; \hat{\sigma}_{ik}, \hat{\delta} \right)dz.
\label{MarginalDensity_AGEB} 
\end{equation}
where $i$,$j$ are the respective residence and visiting patches ($i,j =1,\ldots,n$).

As we mention above, the integral is numerically solved using uniform sampling Monte Carlo integration. For this we randomly  sampled 2-dimensional points within the area of Hermosillo and its metropolitan area, identified the patch (AGEB) to which each belongs and evaluate the density $h_{ik} \left(z; \hat{\sigma}_{ik}, \hat{\delta} \right)$ on them. 


The result of this analysis renders 6 different matrices, each  of dimension  close to $502 \times 502$, as we consider $n=502$ different patches (AGEBs) and few of them did not keep any IDs after filtering. The complexity of this level of information makes it challenging to convey solely through figures. A map may only capture a fraction of this data, such as a single row of the ROM, depicting the expected proportion of time that individuals residing in an AGEB would spend in all other AGEBs. 

Given the inherent complexity of representing the entire residence-mobility matrix in relation to spatial layout, we utilize alternative metrics to effectively identify certain attributes of mobility patterns relative to the AGEBs of residency. The first approach involves assessing the distance between the ROMs computed from both parts of each period. The second alternative entails presenting two maps that illustrate differences in two specific aspects of mobility: namely, the absence of mobility and the proportion of time spent by all individuals (who leave their AGEB at any given time point) in other AGEBs

\subsubsection{Matrix distances}

As our objective of assessing differences between the estimated matrices by period, we compute three of the most relevant matrix distances for the part $A$ and $B$ matrices in each period. These distance functions are the  Manhattan (Mh), Euclidean (E) and Minkowski (Mk) distances, which can be computed for any two matrices with the same dimensions. Let $[C]_{ij}=c_{ij}$ and $[D]_{ij}=d_{ij}$ be two real matrices with the same dimensions. Then Minkowski distances is defined as follows:
$$
    d_{\text{Mk}}(C,D)=\left(\sum_j\sum_i \vert c_{ij}-d_{ij}\vert^p\right)^{1/p}, \ \ \ p\in \mathbb{N}.
$$

The Manhattan and Euclidean distances correspond, respectively, to the Minkowski distance with $p=1$ and $p=2$. We can be observed that these functions are symmetric and null only when all the entries of the matrices coincide.

The ROMS are estimated using the filtered data in each period-part and the number of residents may differ from one to the other. To produce matrices with the same dimensions in both parts of each period, and obtain their distance, we selected the AGEBs with at least one resident in both parts. Then for each period, we compare matrices with number of rows: 469, 479 and 463.

Table~\ref{table:distances} presents the distances for the estimated ROMs derived from the two parts in each period. 

\begin{table}
\centering
\begin{tabular}{@{}lccc@{}}
\toprule
\textbf{} & \textbf{Manhattan} & \textbf{Euclidean} & \textbf{Minkoswki }($p=3$)\\ \midrule
First period ($P1$)  & 220.3977  & 3.0169  & 1.1631 \\
Second period ($P2$) & 196.1594  & 2.3036  & 0.8797\\
Third period ($P3$)  & 182.3303  & 2.0883  & 0.7590\\
 \bottomrule
\end{tabular}
\caption{Matrices distances for estimated matrices in each period.}
\label{table:distances}
\end{table}

We observe that for $P1$, distances are consistently higher. Although its parts are defined when Hermosillo presented a low number of cases (see Figure \ref{fig:Figure_SelectionPeriodsResidenceMatrix}), $P1_A$ occurs several weeks after the first COVID-19 wave, while $P1_B$ precedes it and encompasses two significant festivities: \textit{Halloween} (October 31st) and \textit{Día de Muertos} (November 1st and 2nd). The preparations for these celebrations and the ensuing social gatherings (in homes, markets, and cemeteries) can be associated with the increase in urban mobility and social interaction that facilitated the subsequent wave.

In contrast, the distance between the estimated ROMs for $P2_A$ and $P2_B$ is closer, not because they are closer in time, but due to the observed increase in data and the eventual shift to the orange alert level, which could once again restrict urban mobility.

\subsubsection{Differences for mobility characteristics}

To further study the pattern changes that occurred in Hermosillo, we propose dividing the population into two categories and them estimating the ROM for one of them. These categories are: individuals who never leave their residence patch in each period-time and those who do it. 

We estimate the proportion of individual who never leave their AGEB as the proportion of IDs, in each period-time, that did not have any GPS report outside their AGEB of residence. This information is stored in 6 vectors, each of length close to 500 (as again, few AGEBs did not report any residents).

\begin{figure}[htb]
    \centering
    \begin{subfigure}{\textwidth}
    \centering
        \includegraphics[width=0.95\linewidth]{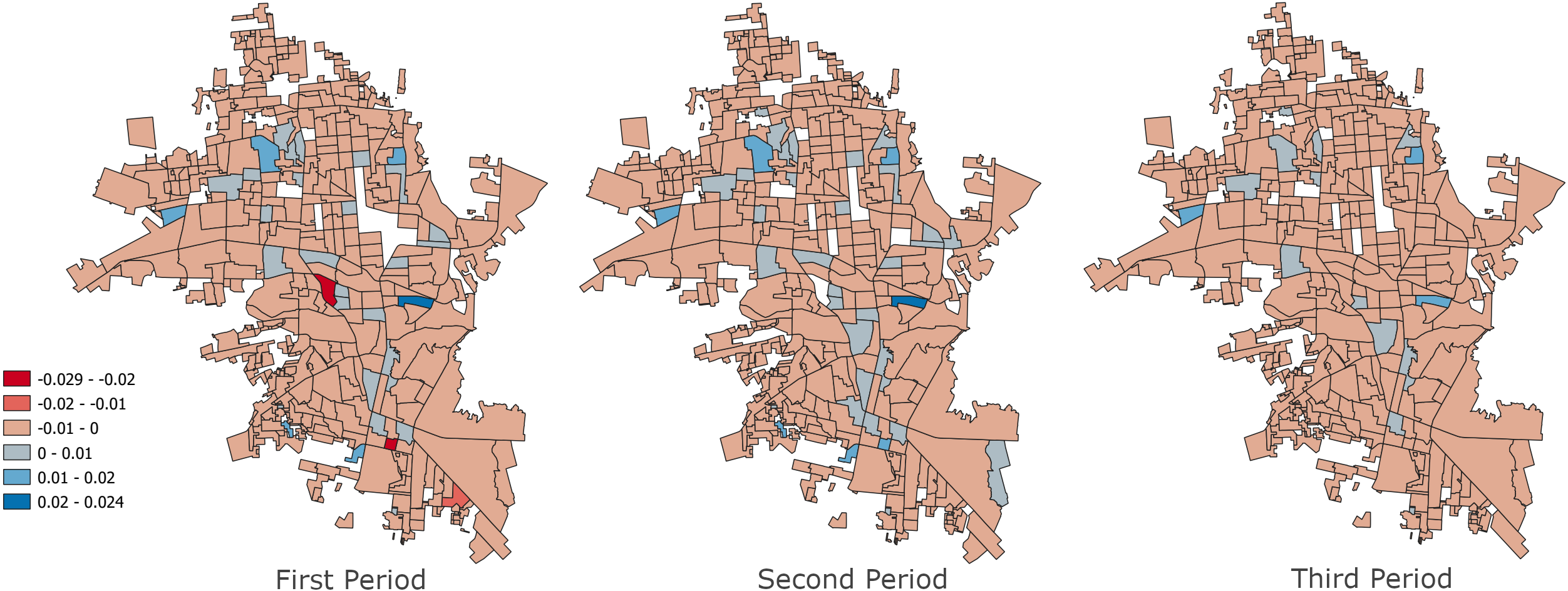}
    \caption{Differences for the proportion of individual that do not leave their AGEB of residency.}
    \label{fig:Diff_alphas}
    \end{subfigure}
    \hfill
    \begin{subfigure}{\textwidth}
    \centering
        \includegraphics[width=0.95\linewidth]{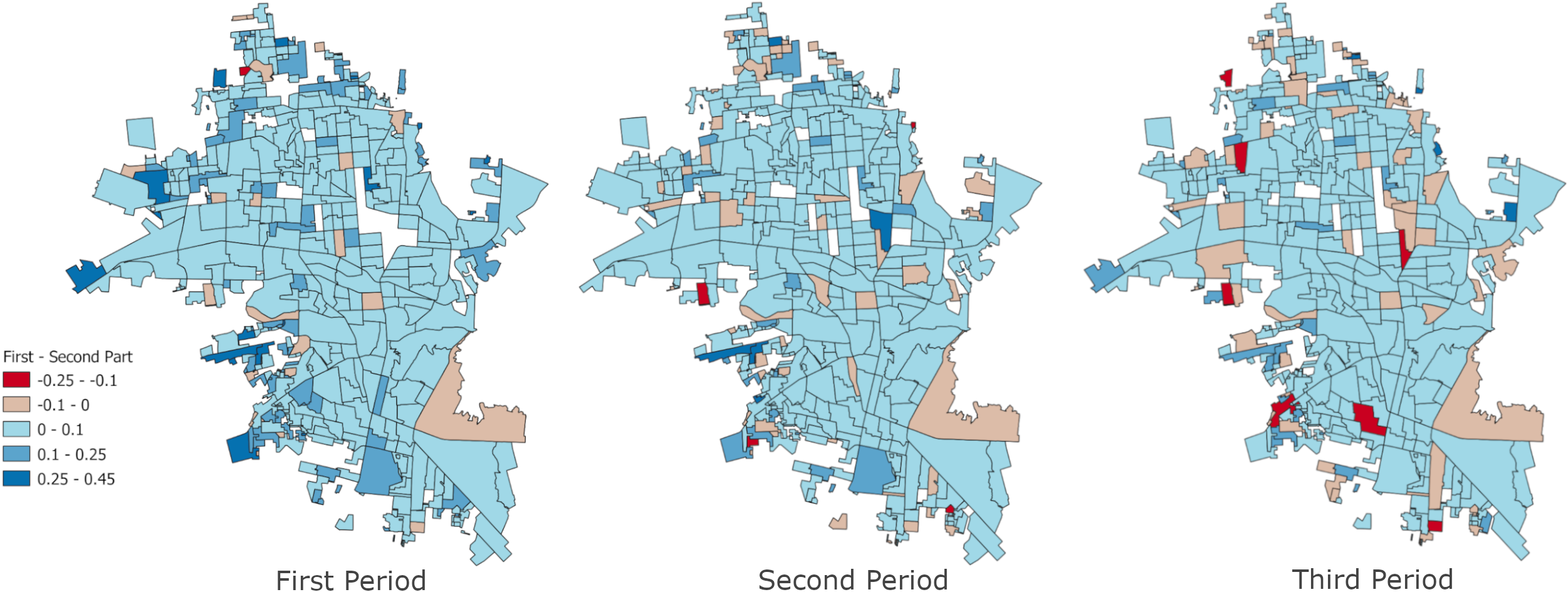}
    \caption{Differences for fraction of time spent in own AGEB of residency for individuals who leave.}
    \label{fig:Diagonal_Difference}
    \end{subfigure}
    \caption{Differences for the proportion of individual that do not leave their AGEBs and times spent within it, for those who leave, between the two parts in each period.}
    \label{fig:differences}
\end{figure}

Considering the AGEBs present in each pair of periods, Figure~\ref{fig:Diff_alphas} illustrates the differences in these proportions (proportion during $Pj_A - $ proportion during $Pj_B$, $j=1,2,3$). 
For each of the three periods, it is evident that  during its second parts, this proportion is  slightly larger for most AGEBs, indicating that there is no evidence that more people tend to leave their own AGEB.

Now, focusing on the data from residents who left their AGEB, Figure~\ref{fig:Diagonal_Difference} depicts the discrepancies in the daily fraction of time they spend in their own AGEB (the proportion of time in their own AGEB during $Pj_A - $ the proportion of time in their own AGEB during $Pj_B$, $j=1,2,3$). This is akin to examining the difference between the diagonals of the square ROMs, which are estimated considering only those individuals with at least one ping outside of their residence. 

From Figures \ref{fig:Diff_alphas} and \ref{fig:Diagonal_Difference} we observe that while the number of indivuals leaving their AGEBs do not increase,  those who do leave, spend less time within their residence. This trend is particularly evident during $P1$ and may potentially be linked to the preparation for festivities, which is also reflected in the larger distance measures between the ROMs for $P1_A$ and $P1_B$ (Table \ref{table:distances}).


The selection of periods and parts was carried out with the objective of testing the consistency and sensitivity of the produced estimates for mobility. As anticipated, for $P3_A$ and $P3_B$, which are consecutive and fall under the yellow epidemic traffic light, the estimates yielded minor differences.

For $P1$ and $P2$, their second parts are separated by gaps of three and four weeks, respectively, from their first parts. As mentioned earlier, the mobility changes observed during $P1$ can be significantly attributed to preparations for and festivities surrounding the holidays during $P1_B$. However,  $P2_B$ includes only the two days of the second festivity (November 1st and 2nd) and the local government decided to increase mobility restrictions to the orange light level on November 9th, resulting in smaller distances for the matrices estimated for $P2$ (see Table \ref{table:distances}).

Regarding the AGEBs and their population, it is notable that AGEBs with the greatest change in mobility for $P1$ and $P2$ are identified as having higher population density and a higher index of marginalization than the rest of the city. These areas correspond to the northern zone of the city (north of Blvd. Progreso) and the southwest zone (the area delimited by Paseo del Río Sonora and Blvd. Las Quintas), which is a residential area with a history of vehicular traffic problems due to the high population density \cite{implan}.

An important aspect to note is that both parts of $P1$ and $P2$ occurred during the yellow epidemic traffic light, and throughout these weeks, the number of reported cases remained relatively stable (see Figure \ref{fig:Figure_SelectionPeriodsResidenceMatrix}). However, from just these two pieces of information (implemented mobility restrictions and recent observed number of cases), we cannot predict the next COVID-19 wave in the city. From this, it becomes evident the practical relevance of harnessing mobility information for predicting and mitigating epidemic outbreaks in a given region. Having this information can help in future planning, as the risk of experiencing a second wave could have been associated with the observed longer visiting times to other AGEBs during $P1_B$ (Figure \ref{fig:differences}).

The mobility patterns depicted in the maps of Figure \ref{fig:differences} represent just a partial visualization of a complex information system that incorporates population density (Figure \ref{fig:AGEB}) and the visiting time patterns of individuals. All this information is included in the epidemic model in the next section and is necessary for justifiable analyses of the effects of mobility on infection cases.


\subsection{The multi-patch epidemic model with mobility and residency}

Human mobility plays major role in the geography of health and epidemiology, as it is a capital factor in the reappearance and persistence of diseases \cite{lee2015role}. Particularly, the explosive urban growth and mobility of people within and between urban regions are factors that affect the geographical dispersion of infectious diseases \cite{amouch2021modeling}. An approach to incorporate the spatial dynamics of mobility into infectious diseases is the use of mathematical models with differential equations that incorporate residency and mobility  to describe human mobility within and between geographic regions in a synthesized manner \cite{driessche2008spatial, herrera2011multiple, khan2009spread, akuno2023multi}. To our knowledge, articles following this approach to modeling disease dynamics focus on specifying particular structures for populaton residence, mobility or occupation. They implement human mobility models that are often not estimated (such as gravity and  radiation models),  or  theoretically or through simulations these works demonstrate the mobility effects on various characteristics of the studied disease dynamics, such as disease transmission, endemicity, and  epidemic peak and duration. However, from an applied standpoint, an important challenge lies in estimating the occupation time by residence. In fact, the problem of estimating ROMs in urban areas poses a theoretical and computational challenge due to the heterogeneity of individual human mobility within cities. Simply put, knowing how much time individuals residing in patch $i$ of a city spend in patch $j$ of the same city requires knowledge of which city residents live in patch $i$, as well as their continuous locations over time. The available information systems that come closest to providing continuous location data over time for general population are mobile phone detection systems. However, most of applications using this data estimate only aggregated data such as origen-destination for commuters or migrants, 
and not the fraction of time that we can expect residents spend in each patch. That is, to our knowledge, no formal methodologies have been published regarding the use of mobile phone detection data to estimate the expected time individuals visit each patch given his/her residence location.

\subsubsection{Epidemic model}

In this research, we employ the presented methodology to estimate the mobility and occupation parameters in a multi-patch epidemic model and assess their effect on it. To achieve this, we utilize the parameters estimated during the different periods as outlined in Table~\ref{table:periods_definitions}.

The following meta-population multi-patched SEIRS compartmental model has been theoretically studied in \cite{akuno2023multi} and we refer the reader to this article to have the intuition behind its formulation and derivation. 

The considered epidemic model evolves in $n$ different patches with $\{N_i\}_{i=1}^n$ inhabitants, and it is formally posed as:
\begin{align}
\begin{cases}
\dot{S}_{i} & =\ \Lambda_{i}-\beta_{i}(1-\alpha_i)S_i \frac{(1-\alpha_i)I_i+\sum_{k=1}^n \tilde{p}_{k i} I_{k}}{(1-\alpha_i)N_i+\sum_{k=1}^n \tilde{p}_{k i}N_{k}} -\sum_{j=1}^n\left(\beta_{j} \tilde{p}_{i j} S_{i}  \frac{(1-\alpha_j)I_j+\sum_{k=1}^n \tilde{p}_{k j} I_{k}}{(1-\alpha_j)N_j+\sum_{k=1}^n \tilde{p}_{k j} N_{k}}\right) \\ &- \mu_{i}S_{i}+\tau_{i}R_{i} \\
\dot{E}_{i} & =\ \beta_{i}(1-\alpha_i)S_i \frac{(1-\alpha_i)I_i+\sum_{k=1}^n \tilde{p}_{k i} I_{k}}{(1-\alpha_i)N_i+\sum_{k=1}^n \tilde{p}_{k i} N_{k}} +\sum_{j=1}^n\left(\beta_{j} \tilde{p}_{i j} S_{i}  \frac{(1-\alpha_j)I_j+\sum_{k=1}^n \tilde{p}_{k j} I_{k}}{(1-\alpha_j)N_j+\sum_{k=1}^n \tilde{p}_{k j} N_k}\right)\\ & - (\kappa_i+\mu_i) E_i \\
\dot{I}_{i} & =\kappa_i E_{i} - (\gamma_i+\psi_i+\mu_{i})I_{i} \\
\dot{R}_{i} & =\ \gamma_i I_i - (\tau_i + \mu_i)R_i
\end{cases}
\label{seirs_system}
\end{align}
for $i=1,\ldots,n$ and where $i$ and $j$ are the index for any two patches. The term $\tilde{p}_{k j}$ corresponds to the parameter product $\alpha_kp_{k j}$, $i=1,2,\cdots,n$. 

Table \ref{table:nonlin} contains the definition of the parameters used in the model.The parameters $\boldsymbol{\alpha}=\{\alpha_i\}_{i=1}^n$ and $\mathbb{P}=\{p_{ij}\}_{i,j=1}^n$ describe the mobility and occupation by residence, and these can be estimated using the methodology used to estimate the ROM. 

\begin{table}
    \centering
    \resizebox{\columnwidth}{!}{
    \centering
    \begin{tabular}{l c l}
        \toprule
        &\textbf{Parameters} & \textbf{Description}\\ [0.5ex]         \midrule
        \multicolumn{2}{l}{Mobility Parameters:} \\
    	&$\alpha_{i}$ & The proportion of individuals that leave their residence patch $i$. \\[2pt]
        &$p_{ij}$ & The proportion of time that an individual from patch $i$ spends\\ 
        &&in patch $j$, given that is one individual that leaves its residence patch.  \\[5pt]
        \multicolumn{2}{l}{Epidemiological Parameters:}\\        
        &$\Lambda_{i}$ & Rage of recruitment of Susceptible individuals in Patch $i$.\\
        &$\beta_{j}$ & Transmission rate for contacts occurring in Patch $j$.  \\
        &$\mu_i$ & Per capita natural death rate in Patch $i$. \\
        &$\gamma_i$ & Per capita recovery rate of individuals in Patch $i$. \\
        &$\tau_i$ & Per capita loss of immunity rate.\\
        &$\psi_i$ & Per capita disease induced death rate of Patch $i$.\\
        &$\kappa_i$ & Per capita rate at which the exposed individuals in patch $i$ becomes infectious.\\
        [1ex]
        \bottomrule
    \end{tabular}
    }
    \caption{Description of the parameters in model (\ref{seirs_system}).}
    \label{table:nonlin}
\end{table}

To appraise the role of mobility on the infection dynamics in Hermosillo, we first estimate the parameters $\boldsymbol{\alpha}$ and $\mathbb{P}$ using the GPS information during each of the periods and parts described in Table \ref{table:periods_definitions}. Then we obtain the solution for the system (\ref{table:nonlin}) for $t\in(0,200)$, using fixed epidemic parameters $\Lambda_{i}$,  $\beta_{j}$, $\mu_i$, $\tau_i$, $\psi_i$ and $\kappa_i$, $i=1,\ldots,n$, and each of the estimated mobility parameters $\boldsymbol{\alpha}$ and $\mathbb{P}$. Finally, we compare the effects of changes in the the resulting mobility parameter by obtaining the difference between the two generated infection counts for each period.

For instance, to measure the epidemic effect of mobility modification that occurred during $P1$,  we estimate the mobility parameters for $P1_A$ and $P1_B$, and compute the difference between the infection counts from  (\ref{table:nonlin}) and under each residence and  mobility matrix. 

\subsubsection{Estimates for mobility parameters}

The mobility parameters categorize the population into two groups: individuals who never leave their residential patch and those who do. The parameter $\boldsymbol{\alpha}$ describes the distribution for the proportion of individuals who leave their patch, while $\mathbb{P}$ refers to the occupation times of individuals in this same category.

We estimate the parameter $\boldsymbol{\alpha}=\{\alpha_i\}_{i=1}^n$ in a similar way to how we produced the vectors depicted in Figure~\ref{fig:Diff_alphas}. For each period-part and $i$, we estimate $\alpha_i$ as the proportion of IDs inhabiting AGEB $i$ and registering at least one ping outside AGEB $i$. Then, the vectors represented in Figure~\ref{fig:Diff_alphas} correspond to the estimation of $1-\alpha_i$, as the percentages are obtained relative to the total.

The matrix $\mathbb{P}=\{p_{ij}\}_{i,j=1}^n$ corresponds to the ROM for individuals who leave their patch. We estimate it as the ROM but considering only IDs that where used in the computation of $\boldsymbol{\alpha}$. In other words,  we filter the data to retain only IDs that registered at least one ping outside their AGEB of residence and use (\ref{MarginalDensity_AGEB}). Then  $\mathbb{P}$ is the ROM  conditioned to individuals leaving their AGEB.

The epidemic model does not consider a ROM for individuals not leaving their patch, as it certain that they spend all day withing their own patch.

\subsubsection{Mobility effects on the outbreaks}

In order to examine how mobility influences the spread of disease across all the AGEBs in Hermosillo, we set the epidemic parameters and initial conditions for all simulations. The initial number of infected individuals was determined based on observed COVID-19 infection data in Hermosillo, with one initial case each in four AGEBs identified by their respective IDs: 2956, 3367, 5734, and 6200. Thus, for the numerical simulation, the initial values were set as follows:
$$
E_{2956}(0) = E_{3367}(0) = E_{5734}(0) = E_{6200}(0) = 1, \ E_{i}(0)=0,  \text{ for } i\notin \{2956, 3367, 5734, 6200\}
$$
$$
I_{2956}(0) = I_{3367}(0) = I_{5734}(0) = I_{6200}(0) = 1, \ I_{i}(0)=0, \text{ for } i\notin \{2956, 3367, 5734, 6200\}
$$
$$
R_i(0)=0, \text{ and } S_i(0)=N_i - (E_i(0)+I_i(0)+R_i(0)) \text { for } i\in \{1,2,\cdots, n\}.
$$

The demographic parameters used in the simulation were obtained from existing literature. The crude death rate for the state of Sonora, Mexico, according to \cite{knoema2021}, is 6\% per 1,000 persons per year. As we are modeling infection dynamics on a daily basis, we used a constant natural death rate of $\mu = 0.06/(1000\times 365)$ for each individual in any AGEB. Population sizes, $N_{i}$, for each AGEB were obtained from the 2019 Mexico census conducted by the National Institute of Statistics and Geography (INEGI).

For this analysis, we assumed an almost constant population size. Utilizing the natural crude death rate 
we calculated the recruitment rate of susceptibles for each AGEB as $\Lambda_{i} = \mu N_{i}$. Following \cite{lin2020conceptual}, who posited that the contact rate of COVID-19 falls within the range $\beta\in(0.5944, 1.68)$, we set $\beta_{i} = 1.5 $ as the daily rate of infection in each AGEB. Latent and recovery rates, as well as the rate at which recovered individuals become susceptible, were set as $\kappa_{i}^{-1} = 7$ days, $\lambda_{i}^{-1} = 14$ days, and $\tau_{i}^{-1} = 180$ days, respectively, based on diverse literature on COVID-19.

\begin{figure}[htb]
    \centering
    \begin{subfigure}{0.5\textwidth}
        \centering
        \includegraphics[width=0.9\linewidth]{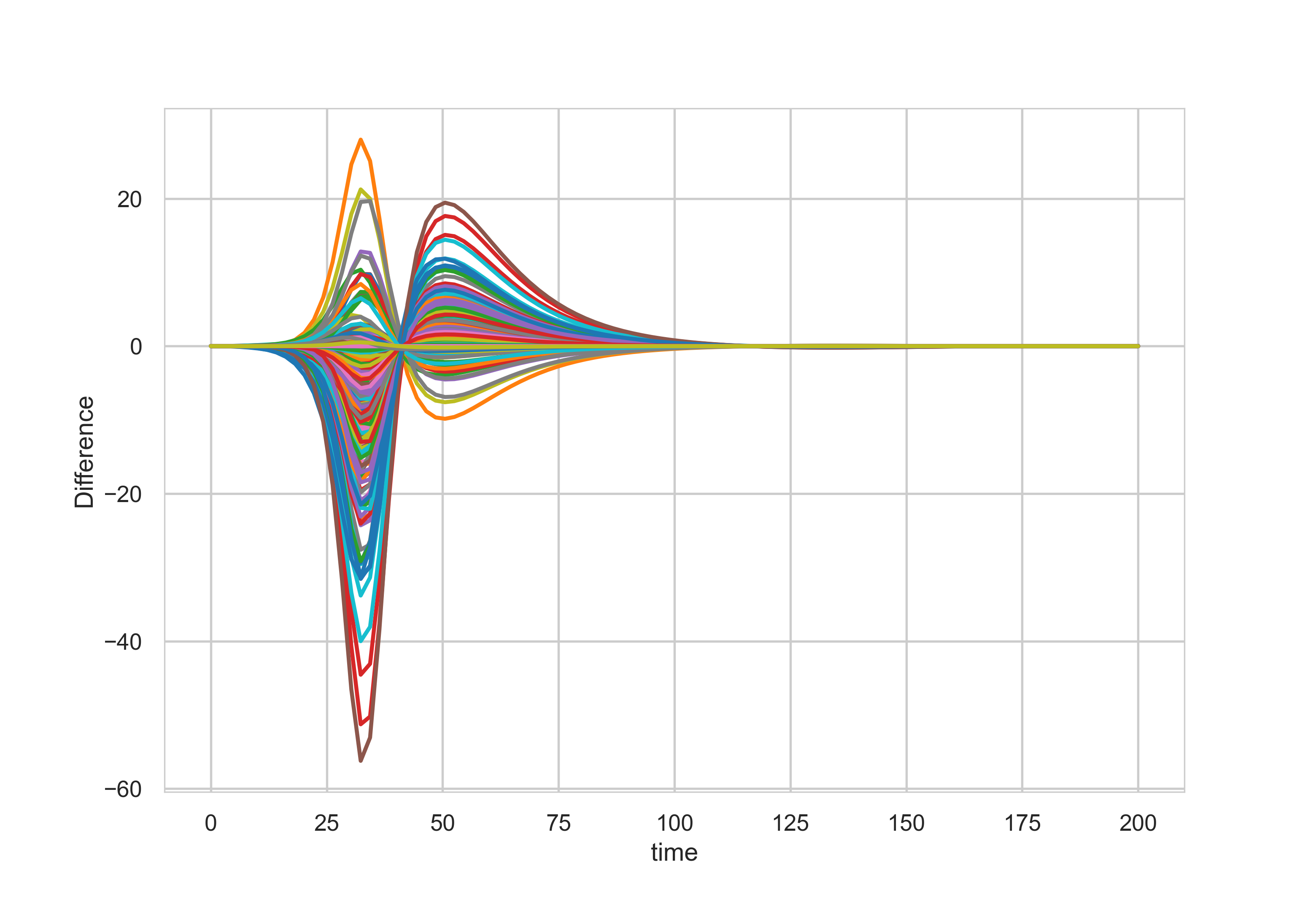}
        \subcaption{$P1_A$ and $P1_B$.}
        \label{difference_FP_FP_and_FP_SP}
    \end{subfigure}\hfill
    \begin{subfigure}{0.5\textwidth}
        \centering
        \includegraphics[width=0.9\linewidth]{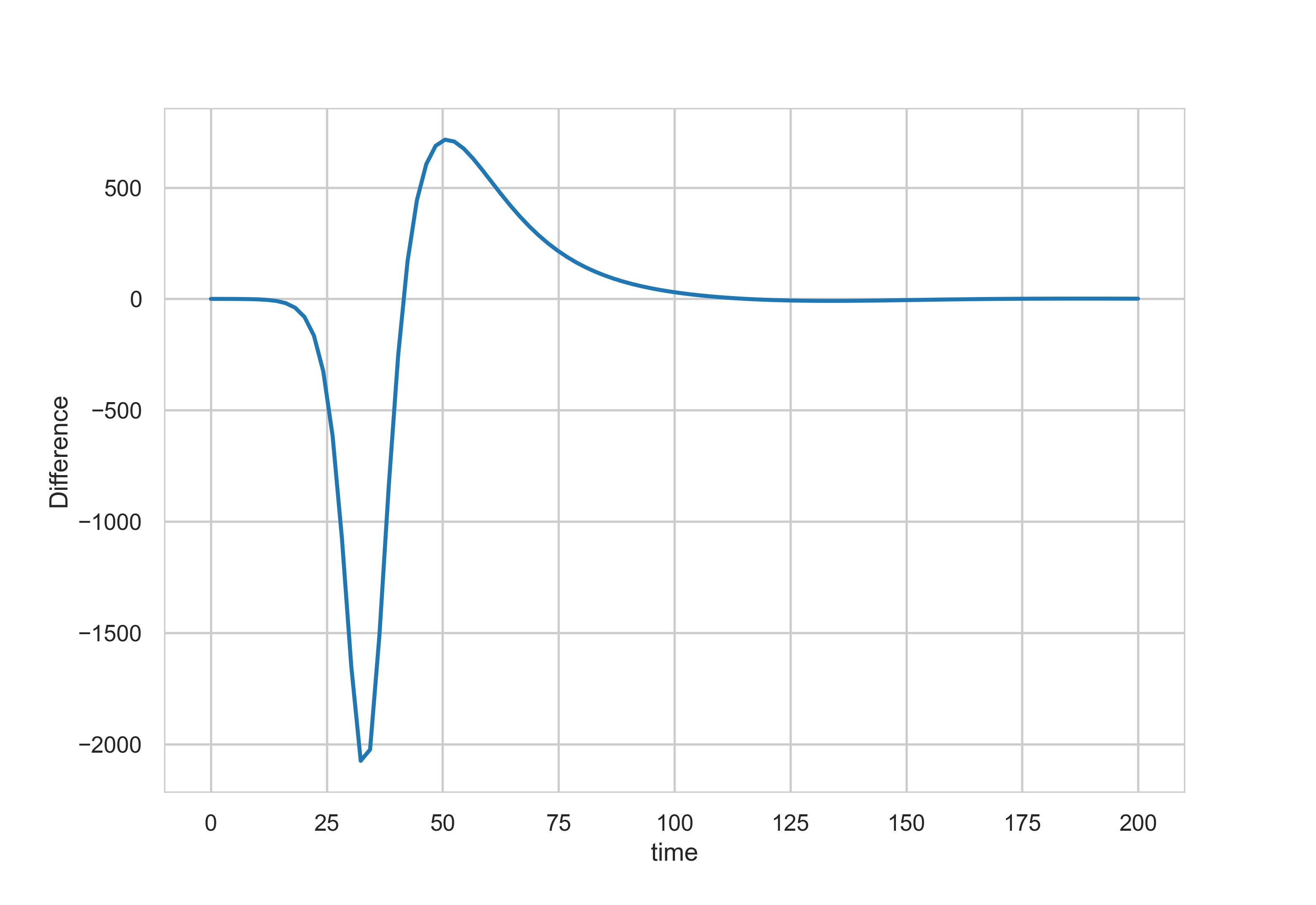}
        \subcaption{$P1_A$ and $P1_B$.}
        \label{difference_global_FP_FP_and_FP_SP}
    \end{subfigure}\hfill
 
  \begin{subfigure}{0.5\textwidth}
        \centering
        \includegraphics[width=0.9\linewidth]{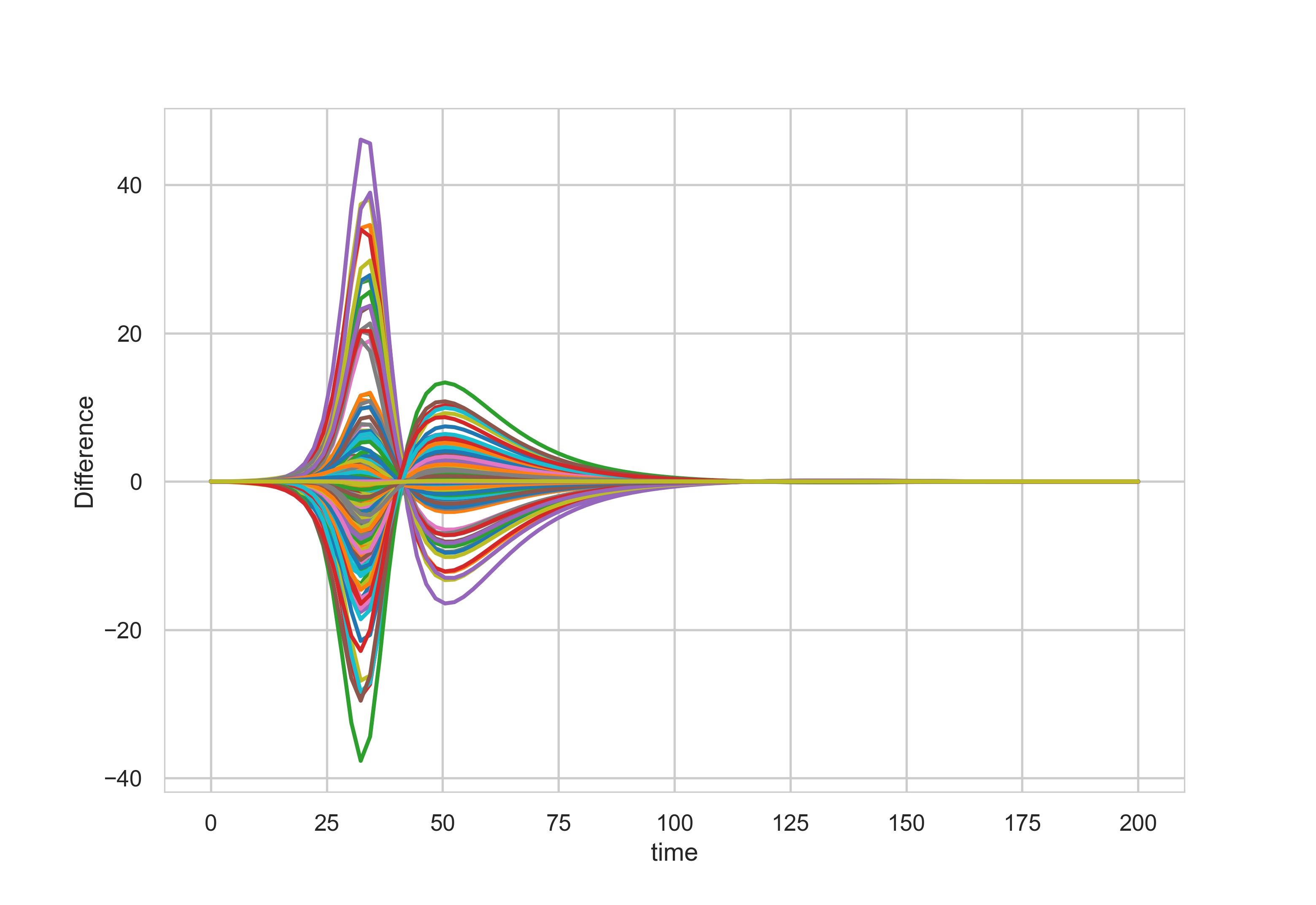}
        \subcaption{$P2_A$ and $P2_B$.}
        \label{difference_SP_FP_and_SP_SP}
    \end{subfigure}\hfill
    \begin{subfigure}{0.5\textwidth}
        \centering
        \includegraphics[width=0.9\linewidth]{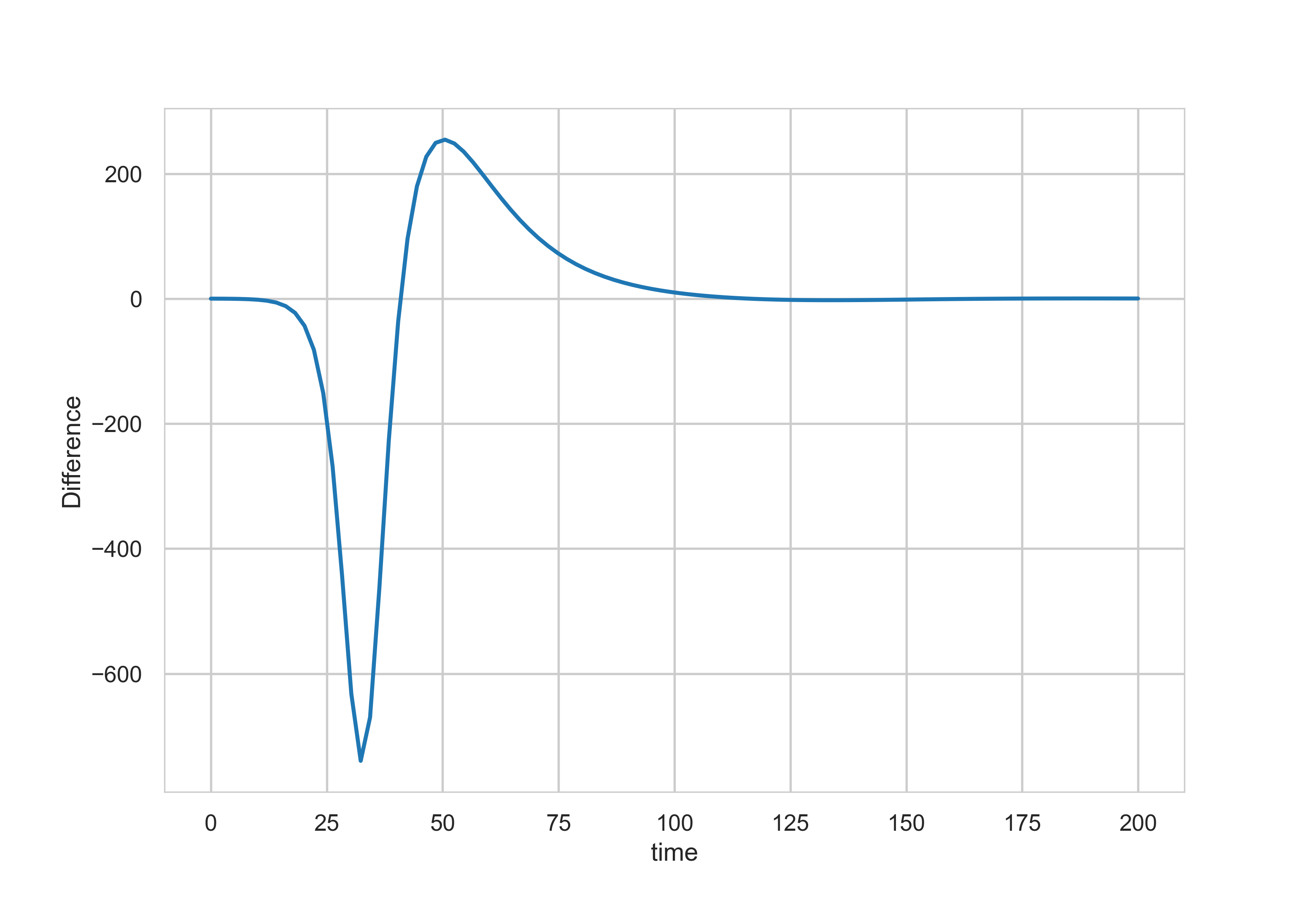}
        \subcaption{$P2_A$ and  $P2_B$.}
        \label{difference_global_SP_FP_and_SP_SP}
    \end{subfigure}\hfill

    \begin{subfigure}{0.5\textwidth}
        \centering
        \includegraphics[width=0.9\linewidth]{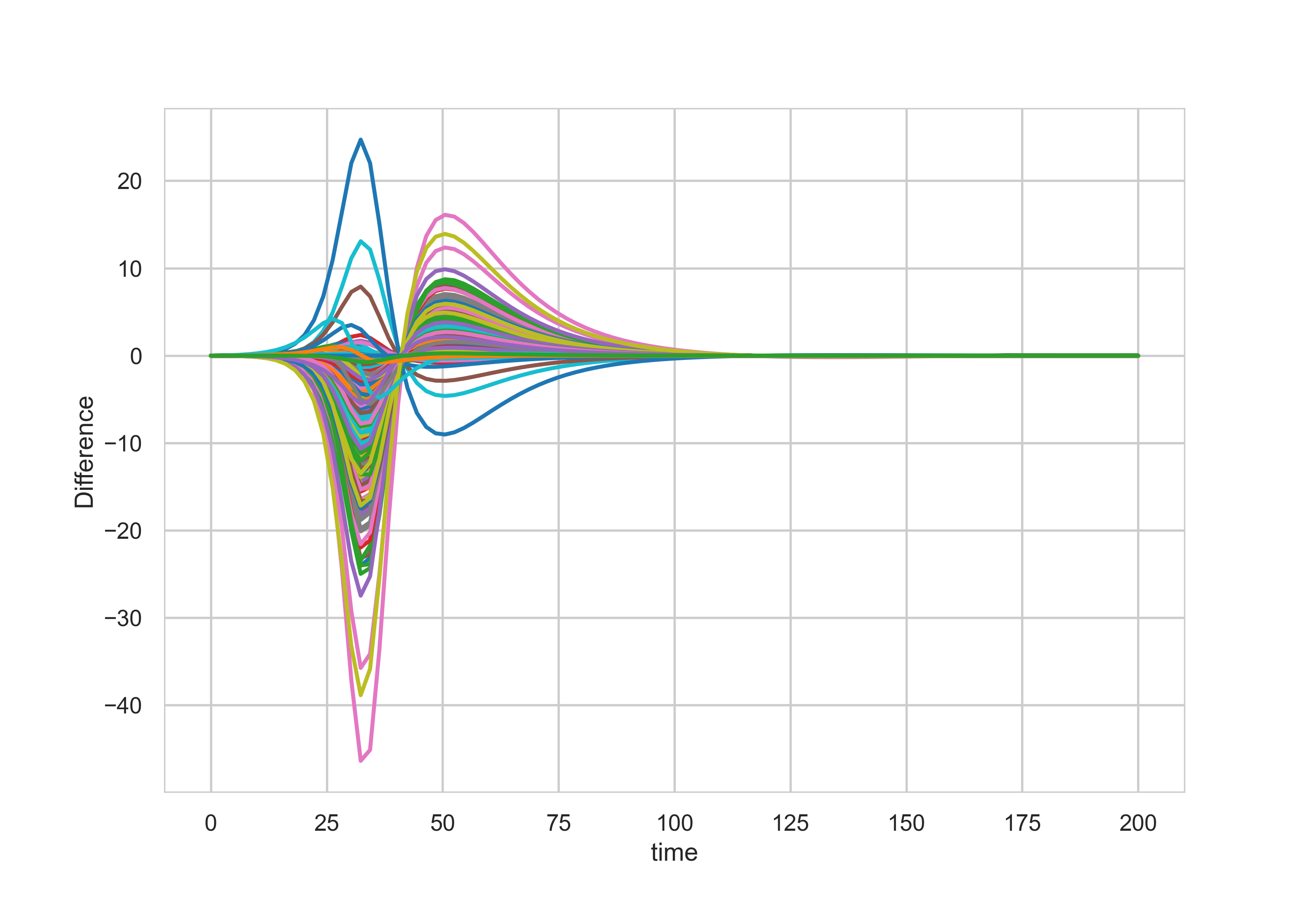}
        \subcaption{$P3_A$ and $P3_B$.}
        \label{difference_TP_FP_and_TP_SP}
    \end{subfigure}\hfill
    \begin{subfigure}{0.5\textwidth}
        \centering
        \includegraphics[width=0.9\linewidth]{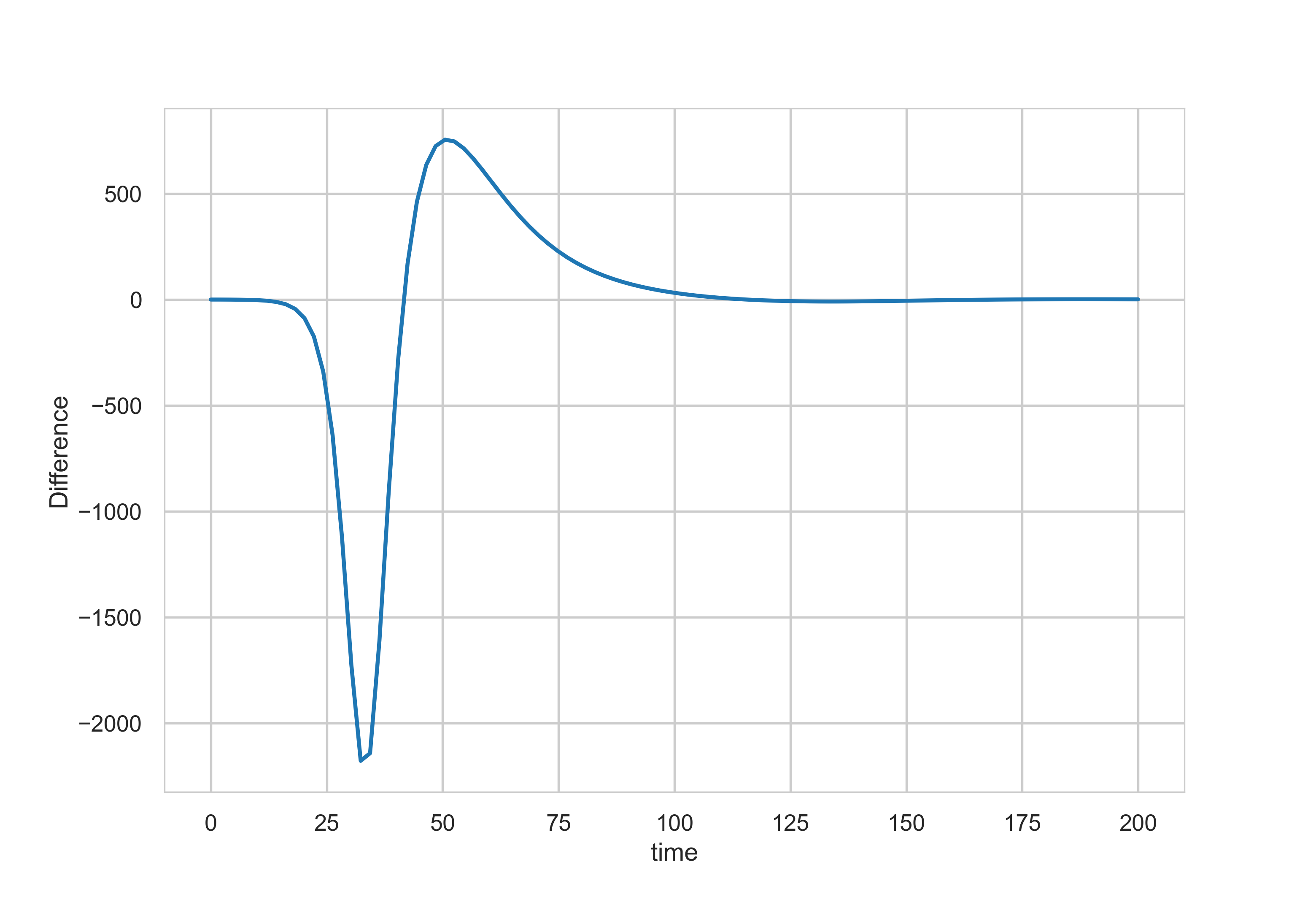}
        \subcaption{$P3_A$ and $P3_B$.}
        \label{difference_global_TP_FP_and_TP_SP}
    \end{subfigure}\hfill
        
    \caption{Differences in individual AGEBS (left) and global (right) infection curves and proportions for the first and second parts of each period.}
    \label{All_FP_and_All_SP}
\end{figure}

Figure \ref{All_FP_and_All_SP} illustrates the disparities in infection counts derived from estimated mobility parameters across different periods (``$I(t)$ under $Pj_A$'' $ - $ ``$I(t)$ under $Pj_B$'', $j=1,2,3$), shedding light on the impact of mobility on disease dynamics both at the level of individual AGEBs and globally. For instance, Figure \ref{difference_FP_FP_and_FP_SP} reveals a swift progression of the disease in most AGEBs during $P1_B$ compared to $P1_A$, particularly within the initial period of approximately $t=30$ days. This suggests that for the majority of AGEBs, their epidemic peak occurred earlier under the estimated mobility parameters in $P1_B$. As the epidemic curve evolves under the estimations derived from $P1_A$, curves in Figure \ref{difference_FP_FP_and_FP_SP} change signs.

Figure \ref{difference_global_FP_FP_and_FP_SP} showcases the disparity in the sum of infection counts for each AGEB between $P1_A$ and $P1_B$ over time. Due to varying population sizes among AGEBs, their contributions to the total number of cases vary. Notably, a few curves deviating from the described pattern do not significantly impact the overall trend at the global level. Similar patterns are discernible in Figures \ref{difference_SP_FP_and_SP_SP} and \ref{difference_global_SP_FP_and_SP_SP}, as well as Figures \ref{difference_TP_FP_and_TP_SP} and \ref{difference_global_TP_FP_and_TP_SP}, illustrating disparities in individual AGEB and global infection counts across $P2$ and $P3$, respectively.

From Table~\ref{table:distances}, we have that $P1$ had the most distant estimated ROM matrices. Nonetheless, Figure~\ref{All_FP_and_All_SP} reveals that the global evolution of diseases from the estimated ROM in $P1$ and $P3$ exhibits similar magnitudes (though both are more pronounced than in the case of $P2$). Indeed, during $P2_B$, the region was under yellow, transitioning to orange epidemic traffic light restrictions by the government, which could realistically reduce infections, as depicted in Figure \ref{difference_global_SP_FP_and_SP_SP}.

The similarity observed in the global differences for the epidemic curves for $P1$ and $P3$ (Figures \ref{difference_global_FP_FP_and_FP_SP} and \ref{difference_global_TP_FP_and_TP_SP}) cannot be directly explained by the distance functions. These functions were obtained for the estimated ROMs and not the estimates of the decoupled mobility information $\boldsymbol{\alpha}$ and $\mathbb{P}$. From Figure \ref{fig:differences} we compare the changes in both parts for $P1$ and $P3$. We observe that in both cases individuals who traveled outside their AGEB spent a large proportion of their time in other AGEBs during their second parts. However, for $P1$ individuals who traveled spent more time outside, and for $P3$ the proportion of individuals leaving is smaller. We have to stress that these figures alone cannot fully explain the similarity in the simulated outbreaks under $P1$ and $P3$, as they represent only a fraction of the categorized mobility information and do not account for other important factors such as AGEB population sizes.

What we can conclude is that epidemic evolution is deeply linked to  population and its mobility patterns. It also becomes evident that simple-summarized mobility models that are introduced into epidemic models may fall short of producing acceptable epidemic forecasts. 

We cannot underestimate the effect of mobility in the epidemic dynamics, but it is linked to complex high-dimensional models that brings numerous challenges when directly addressing its statistical inference or model fitting. To confront this, we divide the estimation problem into two parts: first, the estimation of mobility parameters (that we have presented here), and second, the statistical inference of remaining epidemic parameters. Results for the inference of the epidemic model (\ref{seirs_system}) using this strategy, and based on  COVID-19 incidence data in Hermosillo, are presented in  \cite{akuno2023inference}.


\section{Conclusion}
In this article, we propose a method to incorporate the geolocalization data to estimate not only an origin-destination matrix but more detailed mobility information on the daily visiting time by individuals according to their region of residency. The method is rooted in Brownian-type stochastic models that allow us to obtain a distribution of individual displacement trajectories, and with them, we can answer multiple mobility questions. 

For the considered epidemic model application,  we estimate the ROM at the AGEB level that describes the visiting patterns and times for all the individuals that live in each AGEB.  This matrix, along with the alpha vector, corresponding to the probabilities that any individual move beyond their AGEB,  is an input for a multi-patch SEIR compartmental model applied to the Hermosillo's urban area. 

From the results, it is evident the real impact of population mobility on the epidemiological evolution, as we estimate the residence-mobility matrices utilizing actual data from the city of Hermosillo. 
Consequently, through the epidemic model, we have shown that however small the inter and intra-local residence human mobility may be, the net effect of the same can lead to a noticeable, exigent and momentous local and global infection levels.

As in many models incorporating mobility information, the estimation of residence-mobility parameters can be the first step for fitting or estimating all involved parameters. In the case of epidemic models, we are interested in estimating the parameters and solving questions, such as forecasting and effective intervention identification. 

The one-step estimation of all parameters is likely not feasible, as very complex models prompt non-identifiability problems. Then the produced estimates presented in this research allow us to reduce the model complexity and permit statistical estimation of the infectious agent parameters. This estimation is the subject of ongoing research. 

A challenge that persists in the use of mobile phone data for mobility analysis is the assumption of representativeness of the sample. However, with the increasing use of smartphones, we consider that these databases are becoming more representative samples of the total population. Until then, we can incorporate alternative data to validate or bias-correct some of the estimated parameters.

Furthermore, the selection of the residential AGEB can be enhanced not only by incorporating additional information from diverse sources but also by accounting for the uncertainty associated with the resulting allocation. For more precise results, it is desirable to incorporate this uncertainty into the resulting assertions on the mobility by AGEB of residence. 

Despite the improvements we can make, the mobility and its estimation, using mobile phone data, 
already allows generating important inputs to expand the scope of many mathematical models that describe critical social phenomena such as the economy, violence, transmission of information or infectious diseases. With the proposed methodology, other estimates can also be made that consider, for example, mobility by time ranges, such as nighttime and weekends, or the description of mobility by subgroups, such as mobility by age groups and sex in the area of interest.

\section*{Acknowledgment}
This research has been funded by the CONTEX project 2020-93B Machine-Learning-Assisted Real-Time Simulations and Uncertainty Quantification for Infectious Disease Outbreaks.

The authors would like to express their gratitude to Lumex Consultores S.C for providing the GPS data from mobile phone reports encompassing the city of Hermosillo. This collaboration granted the authors access to the data, under confidentiality conditions, that allows the dissemination of scientific findings derived from this dataset.

\bibliographystyle{siam}
\bibliography{References}

\end{document}